\documentclass{vldb}

\usepackage{graphicx}
\usepackage{balance}  
\usepackage[usenames]{color}
\usepackage[ruled,vlined,linesnumbered]{algorithm2e}

\usepackage{enumitem}
\usepackage{tikz}

\SetNlSty{bfseries}{\color{black}}{}
\let\oldnl\nl
\newcommand{\nonl}{\renewcommand{\nl}{\let\nl\oldnl}}

\newcommand{\tn}[1]{\textnormal{#1}}

\newcommand{\marked}[1]{{\tt marked}[#1]}
\newcommand{\visited}[1]{{\tt visited}[#1]}
\newcommand{\comp}[1]{{\tt comp}(#1)}
\newcommand{\hrc}[1]{{\tt hrc}(#1)}

\newcommand{\mcp}{{\tt ADJ}}
\newcommand{\bmcp}{{\tt binned\_\mcp}}
\newcommand{\pbmcp}[1]{\bmcp[#1]}
\newcommand{\tadd}{\textnormal{add}}

\newcommand{\tvisited}{{\tt visited}~}
\newcommand{\tmarked}{{\tt marked}~}
\newcommand{\tcomp}{{\tt comp}~}
\newcommand{\ahrc}{{\tt hrc}}
\newcommand{\thrc}{{\tt hrc}~}
\newcommand{\trank}{{\tt rank}~}
\newcommand{\tparent}{{\tt parent}~}
\newcommand{\troot}{{\tt root}~}
\newcommand{\tnull}{{\tt null}}
\newcommand{\tsn}{{\tt subnucleus}}
\newcommand{\tsns}{{\tt subnuclei}}

\newcommand{\st}{\text{\textnormal{ s.t. }}}

\newcommand{\mtrue}{\mbox{\textnormal{\textbf{true}}}}
\newcommand{\mfalse}{\mbox{\textnormal{\textbf{false}}}}
\newcommand{\meach}{\text{\textnormal{\textbf{each} }}}
\newcommand{\mand}{\text{\textnormal{ \textbf{and} }}}

\newcommand{\mnot}{\text{\textnormal{\textbf{not} }}}
\DontPrintSemicolon
\newcommand{\kval}{\lambda}
\newtheorem{defn}{Definition}
\newtheorem{corr}{Corollary}

\newcommand{\cS}{\mathcal{S}}

\newcommand{\assk}{{\tt skitter}}
\newcommand{\fbbe}{{\tt Berkeley13}}
\newcommand{\fbmit}{{\tt MIT}}
\newcommand{\fbst}{{\tt Stanford3}}
\newcommand{\fbtx}{{\tt Texas84}}
\newcommand{\twhg}{{\tt twitter-hb}}
\newcommand{\wgo}{{\tt Google}}
\newcommand{\wuk}{{\tt uk-2005}}
\newcommand{\wiki}{{\tt wiki-0611}}

\newcommand{\sassk}{{\tt SK}}
\newcommand{\sfbbe}{{\tt BE}}
\newcommand{\sfbmit}{{\tt MIT}}
\newcommand{\sfbst}{{\tt ST}}
\newcommand{\sfbtx}{{\tt TX}}
\newcommand{\stwhg}{{\tt TW}}
\newcommand{\swgo}{{\tt GO}}
\newcommand{\swuk}{{\tt UK}}
\newcommand{\swiki}{{\tt WK}}

\begin{document}

\title{Fast Hierarchy Construction for Dense Subgraphs}

\numberofauthors{2} 
\author{
\alignauthor
Ahmet Erdem Sar{\i}y\"{u}ce\\
	\affaddr{Sandia National Laboratories}\\
       	\affaddr{Livermore, CA, USA} \\
	\email{asariyu@sandia.gov}
\alignauthor
Ali Pinar\\
	\affaddr{Sandia National Laboratories} \\
	\affaddr{Livermore, CA, USA} \\
	\email{apinar@sandia.gov}
}
\date{30 July 1999}

\maketitle

\begin{abstract}
Discovering dense subgraphs and understanding the relations among them is a fundamental problem in  graph mining. We want to not only identify dense subgraphs, but also build a hierarchy among them (e.g., larger but  sparser subgraphs formed by two smaller dense  subgraphs).   
Peeling algorithms ($k$-core, $k$-truss, and nucleus decomposition) have been effective to locate many dense subgraphs.
However, constructing a hierarchical representation of density structure, even correctly computing the connected $k$-cores and $k$-trusses, have been mostly overlooked. Keeping track of connected components during peeling requires an additional traversal operation, which is as expensive as the peeling process. In this paper, we start with a thorough survey and point to nuances in problem formulations that lead to significant differences in runtimes.  
We then propose efficient and generic algorithms to construct the hierarchy of dense subgraphs for $k$-core, $k$-truss, or any nucleus decomposition.
Our algorithms leverage the disjoint-set forest data structure to efficiently construct the hierarchy during traversal.
Furthermore, we introduce a new idea to avoid traversal.
We construct the subgraphs while visiting neighborhoods in the peeling process, and build the relations to previously constructed subgraphs.
We also consider an existing idea to find the $k$-core hierarchy and adapt for our objectives efficiently.
Experiments on different types of large scale real-world networks show significant speedups over naive algorithms and existing alternatives.
Our algorithms also outperform the hypothetical limits of any possible traversal-based solution.
\end{abstract}

\section{Introduction}\label{sec:intro}

\noindent Graphs are used to model relationships in 
many applications such as  sociology, the WWW, cybersecurity, bioinformatics, and infrastructure.
Although the real-world graphs are sparse ($|E| << |V|^2$), vertex neighborhoods are dense~\cite{Gleich12}.
Clustering coefficients~\cite{WaSt98}, and transitivity~\cite{WaFa94} of real-world networks are also high and suggest the micro-scale dense structures.
Literature is abundant with the benefits of dense subgraph discovery for various applications~\cite{Lee10,Gionis15}.
Examples include finding communities in web~\cite{Kumar99, Dourisboure07}, and social networks~\cite{Huang14}, detecting spam groups in web~\cite{Gibson05}, discovering migration patterns in stock market~\cite{Du09}, improving software understanding by analyzing static structure of large-scale software systems~\cite{Zhang09}, analyzing gene co-expression networks~\cite{Zhang05}, finding DNA motifs~\cite{Fratkin06}, quantifying the significance of proteins~\cite{Wuchty05} and discovering molecular complexes~\cite{Bader03} in protein interaction networks, identifying real-time stories in microblogging websites~\cite{Angel12}, and improving the throughput of social-networking sites~\cite{Gionis13}.
	
$k$-core~\cite{Seidman83, MaBe83}, $k$-truss~\cite{Saito06, Cohen08, Zhang12, Verma12, Colomer13, Zhao13, Huang14}, and their generic variant for larger cliques, nucleus decomposition~\cite{Sariyuce15}, are deterministic algorithms which are effective and efficient solutions to find dense subgraphs and creating hierarchical relations among them.
They also known as \emph{peeling algorithms} due to their iterative nature to reach the densest parts of the graph.
Hierarchy has been shown to be a central organizing principle of complex networks, which is useful to relate communities of a graph and can offer insight into many network phenomena~\cite{Clauset08}.
Peeling algorithms do not aim to find a single optimum dense subgraph, but rather gives many dense subgraphs with varying sizes and densities, and \textbf{\textit{hierarchy}} among them, if supported by a post-processing traversal step~\cite{MaBe83, Sariyuce15}.

\vspace{-2ex}
\subsection{Problem, Misconception and Challenges}
	
\noindent We focus on undirected, unattributed graphs.
Hierarchy of dense subgraphs is represented as the tree structure where each node is a subgraph, each edge shows a containment relation, and the root node is the entire graph.
The aim is to efficiently find the hierarchy by using peeling algorithms.

\textbf{\textit{Misconception in the literature}}: Recent studies on peeling algorithms has interestingly overlooked the connectivity condition of $k$-cores and $k$-trusses.
In the original definition of $k$-core, Seidman states that $k$-core is the maximal and connected subgraph where any vertex has at least degree $k$~\cite{Seidman83}.
However, almost all the recent papers on $k$-core algorithms~\cite{Cheng11, Giatsidis11, Giatsidis13, Bonchi14, OBrien14, Li14, Khaouid15, Wu15, Wen16, Li14} did not mention that $k$-core is a connected subgraph although they cite Seidman's seminal work~\cite{Seidman83}.
On the $k$-truss side, the idea is introduced independently by Saito \emph{et al.}~\cite{Saito06} (as $k$-dense), Cohen~\cite{Cohen08} (as $k$-truss), Zhang and Parthasarathy~\cite{Zhang12} (as triangle $k$-core), and Verma and Butenko~\cite{Verma12} (as $k$-community).
They all define $k$-truss as a subgraph where any edge is involved in at least $k$ triangles.
Regarding the connectivity, Cohen~\cite{Cohen08}, and Verma and Butenko~\cite{Verma12} defined the $k$-truss as a single component subgraph, while others~\cite{Saito06, Zhang12} ignored the connectivity.
In practice, overlooking the connectedness limits the contributions of most previous work regarding the performance and semantic aspects. More details are given in Section~\ref{sec:rel}.

Finding $k$-cores requires traversal on the graph after the peeling process, where maximum $k$-core values of vertices are found.
It is same for $k$-truss and nucleus decompositions where the traversal is done on higher order structures.
Constructing the hierarchy is only possible after that.
However, it is not easy to track nested structure of subgraphs during a single traversal over entire graph.
Traversing $k$-cores is cheap by a simple breadth-first search (BFS) in $O(|E|)$ time.
When it comes to $k$-truss and higher order peeling algorithms, however, traversal becomes much costly due to the larger clique connectivity constraints.

\subsection{Contributions}

\noindent Motivated by the challenging cost of traversals and hierarchy construction, we focus on efficient algorithms to find the $k$-cores, $k$-trusses or any nuclei in general. Our contributions are as follows:
\vspace{-1ex}
\begin{itemize}[leftmargin=*]
\itemsep-0.1em 
\item \textbf{Thorough literature review:} We provide a detailed review of literature on peeling algorithms to point the misconception about $k$-core and $k$-truss definitions.
We highlight the implications of these misunderstandings on the proposed solutions.
We also stress the lack of understanding on the hierarchy construction and show that it is as expensive as the peeling process.

\item \textbf{Hierarchy construction by disjoint-set forest:} We propose to use disjoint-set forest data structure (DF) to track the disconnected substructures that appear in the same node of the hierarchy tree.
Disjoint-set forest is incorporated into the hierarchy tree by selectively processing the subgraphs in a particular order.
We show that our algorithm is generic, i.e., works for \emph{any} peeling algorithm.

\item \textbf{Avoiding traversal:} We introduce a new idea to build the hierarchy without traversal.
In the peeling process, we construct the subgraphs while visiting neighborhoods and bookkeep the relations to previously constructed subgraphs.
Applying a lightweight post-processing operation to those tracked relations gives us all the hierarchy, and it works for \emph{any} peeling algorithm.

\item \textbf{Experimental evaluation:} All the algorithms we proposed are implemented for $k$-core, $k$-truss and $(3,4)$-nucleus decompositions, in which peeling is done on triangles and the four-clique involvements.
Furthermore, we bring out an idea from Matula and Beck's work~\cite{MaBe83}, and adapt and implement it for our needs to solve the $k$-core hierarchy problem more efficiently.
Table~\ref{tab:sum} gives a summary of the speedups we get for each decomposition.
Our $k$-core hierarchy algorithm adaptation outperforms naive baseline by $58$ times on \wuk~graph.
The best $k$-truss and $(3,4)$ algorithms are significantly faster than alternatives.
They also beat the hypothetically best possible algorithm (\textsc{Hypo}) that does traversal to find hierarchy.
It is a striking result to show the benefit of our traversal avoiding idea.
\end{itemize}

\begin{table}[!t]
\vspace{-2ex}
\centering
\linespread{0}\selectfont{}
\caption{\small Speedups with our best algorithms for each decomposition. Starred columns (*) show lower bounds, when the other algorithm did not finish in 2 days or did a partial work. Best $k$-truss and (3,4) algorithms are significantly faster than alternatives, and also more efficient than the hypothetically best possible algorithm (\textnormal{\textsc{Hypo}}) that does traversal to find the hierarchy.}
\linespread{1}\selectfont{}
\renewcommand{\tabcolsep}{2pt}
\vspace*{0.5ex}
\begin{tabular}{|l||r|r|r|r|r|r|}\hline
\multicolumn{1}{|c||}{} & \multicolumn{1}{c|}{$k$-core} & \multicolumn{3}{c|}{$k$-truss} & \multicolumn{2}{c|}{$(3,4)$ nucleus} \\
&\textsc{Naive} & \textsc{Hypo} & \textsc{Naive} & \textsc{TCP*\tiny{\protect\cite{Huang14}}} & \textsc{Hypo} & \textsc{Naive*} \\ \hline \hline
\fbst	&$	25.50	$x&$	1.10	$x&$	12.58	$x&$	3.41	$x&$	1.48	$x&$	1321.89	$x\\ \hline
\twhg	&$	27.89	$x&$	1.33	$x&$	16.24	$x&$	3.27	$x&$	1.78	$x&$	38.96	$x\\ \hline
\wuk	&$	58.02	$x&$	1.68	$x&$	90.50	$x&$	11.07	$x&$	1.24	$x&$	1.98	$x\\ \hline
\end{tabular}
\label{tab:sum}
\end{table}
\vspace*{-4ex}
\section{Preliminaries}\label{sec:prelim}

This section presents building blocks  for our work. 
 
\subsection{Nucleus decomposition}\label{subsec:nucleus}
\noindent Let $G$ be an undirected and simple graph.
We start by quoting the Definitions 1 and 2 from~\cite{Sariyuce15}.
We use $K_r$ to denote an $r$-clique.

\begin{defn} \label{def:cl-path} Let $r < s$ be positive integers and $\cS$ be a set of $K_s$s in $G$.
\vspace{-1ex}
\begin{itemize}
	\itemsep-0.2em 
	\item $K_r(\cS)$ is the set of $K_r$s contained in some $C \in \cS$. \hspace{-2ex}
	\item The number of $C\in \cS$ containing $u \in K_r(\cS)$ is the \emph{$K_s$-degree} of $u$.
	\item Two $K_r$s $u, u'$ are \emph{$K_s$-connected} if there exists a sequence $u = u_1, u_2, \ldots, u_k = u'$ in $K_r(\cS)$
	such that for each $i$, some $K_s~C \in \cS$ contains $u_i \cup u_{i+1}$.
\end{itemize}
\end{defn}
\vspace{-0.6ex}

These definitions are generalizations of the standard vertex degree and connectedness.
Indeed, setting $r=1$ and $s=2$ (so $\cS$ is a set of edges) yields exactly that.
The main definition is as follows. 

\vspace{-1ex}
\begin{defn} \label{def:nucleus} Let $k$, $r$, and $s$ be positive integers such that $r < s$.  A 
\emph{$k$-$(r,s)$ nucleus} is a maximal union $\cS$ of $K_s$s such that:
\vspace{-1ex}
\begin{itemize}
	\itemsep-0.2em 
	\item The $K_s$-degree of any $u \in K_r(\cS)$ is at least $k$.
	\item Any $u, u' \in K_r(\cS)$ are $K_s$-connected.
\end{itemize}
\end{defn}
\vspace{-0.6ex}

\begin{figure}[!h]
\centering
\vspace*{-2ex}
\linespread{0}\selectfont{}
\caption{\small 2-(2,3) and 2-(2,4) nuclei on left and right.}
\linespread{1}\selectfont
\includegraphics[width=0.5\linewidth]{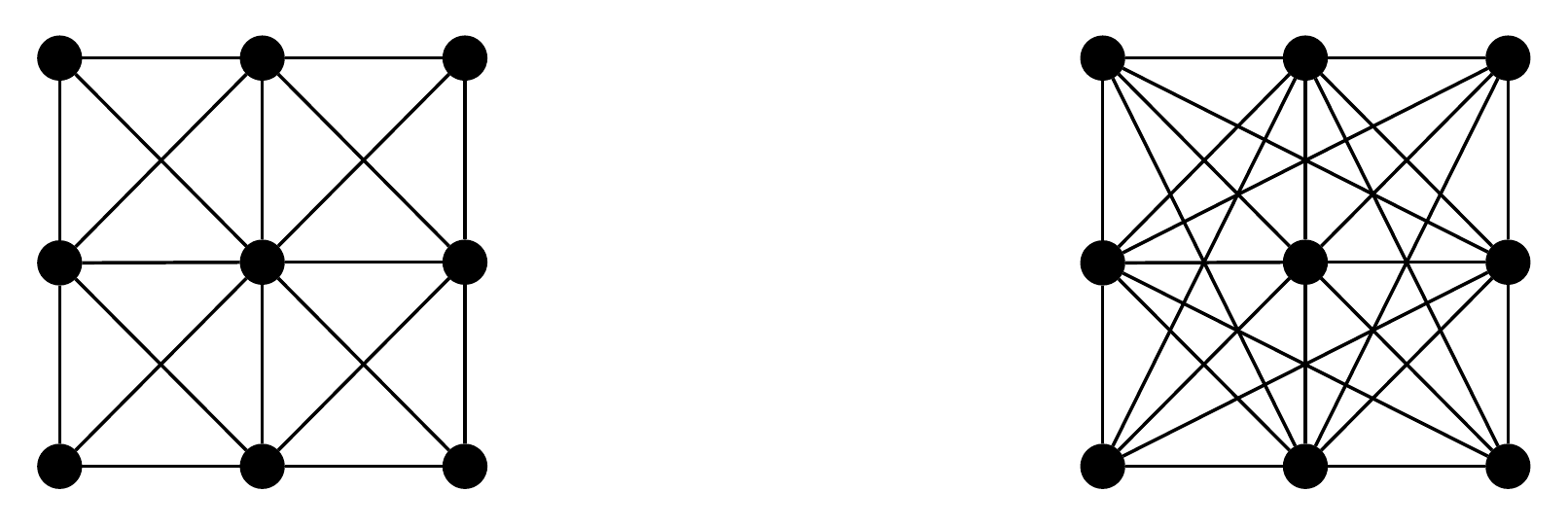}

\label{fig:k2x_example}
\vspace{0.5ex}
\end{figure}

Figure~\ref{fig:k2x_example} gives an example for 2-(2,3) and 2-(2,4) nucleus.
For $r=1, s = 2$, a $k$-(1,2) nucleus is a maximal (induced) connected subgraph with minimum vertex degree $k$.
This is exactly $k$-core~\cite{Seidman83}.
Setting $r = 2, s = 3$ gives maximal subgraphs where every edge participates in at least $k$ triangles, and edges are triangle-connected.
This is almost the same definition of $k$-dense~\cite{Saito06}, $k$-truss~\cite{Cohen08}, triangle-cores~\cite{Zhang12} and $k$-community~\cite{Verma12}.
The difference is on the connectivity condition;~\cite{Cohen08, Verma12} defines $k$-truss and $k$-community as a connected component, whereas~\cite{Saito06, Zhang12} do not mention connectedness, implicitly allowing disconnected subgraphs.
The \emph{$k$-truss community} defined by Huang \emph{et al.}~\cite{Huang14}  is the same as the $k$-(2,3) nucleus: both require any pair of edges to be triangle-connected.
More details can be found in Section~\ref{sec:rel}.
In the rest of the paper, we will be generic and present all our findings for the $k$-$(r,s)$ nucleus decomposition, which subsumes $k$-core and $k$-truss definitions.

For an r-clique $K_r$ $u$, $\omega_{s}(u)$ denotes the $K_s$-degree of $u$.
For a subgraph $H\subseteq G$, $\omega_{r,s}(H)$ is defined as the minimum $K_s$-degree of a $K_r$ in $H$, i.e., $\omega_{r,s}(H)=min\{\omega_s(u): u\in H\}$.

\begin{table}[!t]
\vspace*{-2ex}
\renewcommand{\tabcolsep}{2pt}
\linespread{0}\selectfont{}
\caption{\small Summary of notations}
\linespread{1}\selectfont{}
\begin{tabular}{ | l | l |} \hline
\textbf{Symbol}	&	\textbf{Description} \\\hline\hline
$K_r$			&	$r$-clique; complete graph of $r$ vertices \\ \hline
$\omega_s(u)$		&	$K_s$-degree of $u$; number of $s$-cliques containing $u$ \\ \hline
$\omega_{r,s}(H)$	& 	$min\{\omega_s(u): u\in H\}$; min $K_s$-degree of a $K_r$ in $H$ \\ \hline
$H^u$			&	max $k$-$(r,s)$ nucleus associated with the $K_r$ $u$ \\ \hline
$\kval_s(u)$ 		&	$\omega_{r,s}(H^u)$; max $k$-$(r,s)$ number of the $K_r$ $u$ \\ \hline
$\kval_{r,s}(H)$	&	$min\{\kval_s(u): u\in H\}$; min $\kval_s$ of a $K_r$ in graph $H$ \\ \hline				
$T_{r,s}$		&	sub-$(r,s)$ nucleus; maximal union of $K_r$s of same $\kval_s$  \\ \hline				
\end{tabular}
\vspace{1ex}
\label{tab:notation}
\end{table}

\vspace{-1ex}
\begin{defn} \label{def:max-nucleus}
The \textbf{maximum $\textbf{k}$-$\textbf{(r,s)}$ nucleus} associated with a $K_r$ $u$, denoted by $H^u$, is the $k$-$(r,s)$ nucleus that contains $u$ and has the largest $k=\omega_{r,s}(H^u)$ (i.e., $\nexists~H\st u\in H \,\wedge\, \mbox{$H$ is an $l$-$(r,s)$ nucleus} \,\wedge\, l>k$).

The \emph{maximum} $k$-$(r,s)$ number of the r-clique $u$, denoted by $\kval_s(u)$, is defined as $\kval_s(u) = \omega_{r,s}(H^u)$.
\end{defn}
\vspace{-1ex}
Throughout the paper, $\kval_s(u)$ implies that $u$ is a $K_r$.
We also abuse the notation as $\kval(u)$ when $r$ and $s$ are obvious.
The maximum $k$-$(1,2)$ nucleus is same as the maximum $k$-core, defined in~\cite{Sariyuce13-VLDB}.
For a vertex, $v$, $\kval_2(v)$ is also equal to the maximum $k$-core number of $v$~\cite{Sariyuce13-VLDB}, or core number of $v$~\cite{Cheng11, OBrien14}.
Likewise, for an edge $e$, $\kval_3(e)$ is previously defined as the \emph{trussness} of an edge in~\cite{Huang14}.

Building the $(r,s)$ nucleus decomposition of a graph $G$ is finding the $\kval_s$ of all $K_r$s in $G$ and building the $k$-$(r,s)$ nuclei for all $k$.
The following corollary shows that given the $\kval_s$ of all $K_r$s, all $(r,s)$ nuclei of $G$ can be found.

\vspace{-1ex}
\begin{corr} \label{cor:find-rs}
Given $\kval_s(v)$ for all $K_r$ $v\in G$ and assuming $\kval_s(u) = k$ for a $K_r$ $u$, the maximum $k$-$(r,s)$ nucleus of $u$, denoted by $H^u$, consists of $u$ as well as any $K_r$ $v$ that has $\kval_s(v)\ge k$ and is reachable from $u$ via a path $P$ of $K_s$s such that $\forall_{C\in P},\, \kval_{r,s}(C)\geq k$, where $\kval_{r,s}(C)=min\{\kval_s(u): u\subset C~(u~is~a~K_r)\}$.
\end{corr}
\vspace{-1ex}

Corollary~\ref{cor:find-rs} is easy to see for the $k$-core case, when $r=1, s=2$.
All the traversed vertices are in $H^u$ due to maximality property of $k$-cores, and all the vertices in $H^u$ are traversed due to the connectivity condition, both mentioned in Definition~\ref{def:nucleus}.
For the maximum $k$-$(2,3)$ nucleus, we can also see the equality by Definition~\ref{def:nucleus}.
For all edges $e$, $\kval_3(e)\ge k$ satisfies the first condition, and the path of triangles, which does not contain any edge whose $\kval_3$ is less than $k$, implies the second condition of Definition~\ref{def:nucleus}.

\vspace{-2ex}
\begin{corr} \label{cor:alg-traverse-rs}
$H^u$ can be found by traversing $G$ starting at the $K_r$ $u$ and including each $K_r$ $v$ to $H^{u}$ if
\begin{itemize}
\itemsep-0.1em
\item $\kval_s(v)\ge k$
\item $\exists~a~K_s~C$ s.t. $v\subset C \land \kval_{r,s}(C)\ge k$.
\end{itemize}
Repeating this traversal for all $K_r$s $u\in G$ gives all the $k$-$(r,s)$ nuclei of $G$.
\end{corr}
\vspace{-2ex}

Traversal is trivial for $k$-$(1,2)$ nucleus ($k$-core): include every vertex with greater or equal $\kval_2$.
For $r=2, s=3$, maximum $k$-$(2,3)$ nucleus is found by doing traversal on edges.
Assuming the $\kval_3$ value of the initial edge is $k$, the next edge in the traversal should have the $\kval_3$ value $\geq k$; should be in the same triangle; and all the edges of this triangle should have $\kval_3$s greater-than or equal to $k$.
Similar for $r=3,s=4$; traversal is done on triangles, and neighborhood condition is on the containment of triangles in four-cliques.

\begin{algorithm}[!t]
\small
\caption{\textsc{Set-$\kval$($G, r, s$)}}
\label{alg:set-nucleus}
  Enumerate all $K_r$s in $G(V,E)$\;
  For every $K_r$ $u$, set $\omega_s(u)$ as the number of $K_s$s containing $u$\;
  Mark every $K_r$ as unprocessed\;
  \For{\meach unprocessed $K_r$ $u$ with minimum $\omega_s(u)$}{
  	$\kval_s(u) = \omega_s(u)$, $max\kval=\kval_s(u)$\;
    Find set $\cS$ of $K_s$s containing $u$\label{ln:sk6} \;
    \For{\meach $C \in \cS$}{\label{ln:sk7}
    	\lIf{any $K_r$ $v \subset C$ is processed}{\label{ln:sk8}
    		continue
    	}
    	\For{\meach $K_r$ $v \subset C$, $v \neq u$}{
    		\lIf{$\omega_s(v) > \omega_s(u)$}{
    			$\omega_s(v) = \omega_s(v) - 1$
    		}
    	}
    }
  Mark $u$ as processed\;
  }
  \Return array $\kval_s(\cdot)$ and max$\kval$\;
\end{algorithm}

\begin{algorithm}[!t]
\small
\caption{\textsc{Traversal($G, r, s, \kval(\cdot), max\kval$)}}
\label{alg:traverse-nucleus}  
  \For{\meach $k \in [1, max\kval]$}{
    $\visited{v} = \mfalse, \forall~K_r~v \in G$ \;
    \For{\meach $K_r~u \in G$}{
      \If{$\kval_s(u) = k \mand \mnot\visited{u}$}{
        $Q \leftarrow$ empty queue, $Q$.push($u$)\;
        $S \leftarrow$ empty set, $S$.add($u$)\;
        $\visited{u} \leftarrow \mtrue$ \;
        \While{$\mnot Q$\textnormal{.empty()}}{
          $u \leftarrow Q$.pop()\;        
          \For{\meach $K_r$ $v$ s.t. \\ \label{ln:tr10}
          \nonl $(u\cup v\subset C) \land (C$ is a $K_s) \land (\kval_{r,s}(C) \ge k)$}{
            \If{$\mnot\visited{v}$}{          
              $Q$.push($v$), $S$.add($v$)\;
              $\visited{v} \leftarrow \mtrue$ \;
           }  
         }
       }
       report $S$ \tcp*{output the $k$-$(r,s)$ nucleus}
     }
   }
 }
\end{algorithm}

\begin{algorithm}[!t]
\small
\caption{\textsc{NucleusDecomposition($G, r, s$)}}
\label{alg:naive}
$\kval_s(\cdot)$, max$\kval \leftarrow$ \textsc{Set-$\kval$($G, r, s$)}\tcp{Finding $\kval$s of $K_r$s}
\textsc{Traversal($G, r, s, \kval(\cdot), max\kval$)}\tcp{Finding all $(r,s)$ nuclei}
\end{algorithm}

In summary, $(r,s)$ nucleus decomposition problem has two phases: (1) peeling process which finds the $\kval_s$ values of $K_r$s, (2) traversal on the graph to find all the $k$-$(r,s)$ nuclei.
For $r=1, s=2$ case, the algorithm for finding $\kval_2$ of vertices is based on the following property, as stated in~\cite{MaBe83}: to find the vertices with the $\kval_2$ of $k$, all vertices of degree less than $k$ and their adjacent edges are recursively deleted.
For first phase, we provide the generic peeling algorithm in Alg.~\ref{alg:set-nucleus}, which has been introduced in our earlier work~\cite{Sariyuce15}, and for the second phase, we give the generic traversal algorithm in Alg.~\ref{alg:traverse-nucleus}, which is basically the implementation of Corollary~\ref{cor:alg-traverse-rs}.
The final algorithm, outlined in Alg.~\ref{alg:naive} combines the two.

Lastly, we define sub-$(r,s)$ nucleus and strong $K_s$ -connected ness to find the $K_r$s with same $\kval_s$ values.
We will use them to efficiently locate all the $k$-$(r,s)$ nuclei of given graph.

\vspace{-1ex}
\begin{defn} \label{def:strongly-connected}
Two $K_r$s $u, u'$ with $\kval_s(u) = \kval_s(u')$ are \textbf{strongly $K_s$-connected} if there exists a sequence $u = u_1, u_2$ $, \ldots, u_k = u'$ such that:
\begin{itemize}
\itemsep-0.2em 
\item $\forall i, \kval_s(u_i)=\kval_s(u)$
\item $\forall i, \exists~C$ s.t.
\begin{itemize}
\itemsep-0.1em 
\item $C$ is a $K_s$
\item $(u_i \cup u_{i+1}) \subset C$
\item $\kval_{r,s}(C) = \kval_s(u)$.
\end{itemize}
\end{itemize}
\end{defn}

\begin{defn} \label{def:sub-nucleus}
\textbf{sub-}\tn{(}$\textbf{r,s}$\tn{)} nucleus, denoted by $T_{r,s}$, is a maximal union of $K_r$s s.t.
$\forall~K_r~pair~u, v \subset S$,
\begin{itemize}
\itemsep-0.2em 
\item $\kval_s(u)=\kval_s(v)$
\item $u$ and $v$ are strongly $K_s$-connected
\end{itemize}
\end{defn}
\vspace{-1.5ex}

\noindent The sub-$(1,2)$ nucleus is defined as the subcore in~\cite{Sariyuce13-VLDB, Sariyuce16}. All the notations are given in Table~\ref{tab:notation}.

\subsection{Disjoint sets problem}
\noindent Disjoint-set data structure, also known as union-find, keeps disjoint dynamic sets, and maintains upon the operations that modifies the sets~\cite{CLRS01}.
Each set has a representative.
There are two main operations: \textsc{Union} ($x, y$) merges the dynamic sets with ids $x$ and $y$, and creates a new set, or just merge one of the sets into the other.
\textsc{Find} ($x$) returns the representative of the set which contains $x$.

\LinesNumberedHidden
\begin{algorithm}[!]
\small
\caption{\textsc{Disjoint-Set Forest}}
\label{alg:disjoint-forest}
\underline{\textsc{Link}}($x, y$): \tcp*{$x$ and $y$ are nodes in the tree}
~\lIf{$x$.\tn{rank} $> y$.\tn{rank}}{$y$.\tn{parent} $\leftarrow x$}
~\Else {
~  $x$.\tn{parent} $\leftarrow y$\;
~  \lIf{$x$.\tn{rank} $= y$.\tn{rank}}{$y$.\tn{rank} $\leftarrow y$.\tn{rank} + 1}}
\underline{\textsc{Find}}($x$):\;
~$S \leftarrow$ empty set\;
~\lWhile{$x$.\tn{parent}~is~not~null}{$x \leftarrow x$.parent,  $S$.add($x$)}        
~\lFor{\meach $u \in S$}{$u$.parent $\leftarrow x$}
~\Return x\;
\underline{\textsc{Union}}($x, y$):~\textsc{Link}(\textsc{Find}($x$), \textsc{Find}($y$))
\end{algorithm}
\LinesNumbered

Disjoint-set forest is introduced with two powerful heuristics~\cite{Tarjan75}.
In the disjoint-set forest, each set is a tree, each node in the tree is an element of the set, and the root of each tree is the identifier of that set.
To keep the trees flat, two heuristics are used that complement each other.
First is union-by-rank, which merges the shorter tree under the longer one.
Second heuristic is path-compression that makes each node on the find path point directly to the root.
Time complexity with union-by-rank and path-compression heuristics is $O((m+n)log^*n)$, where $log^*n$ is the inverse Ackermann function which is almost linear~\cite{Tarjan75}. 
Pseudocode for \textsc{Find} and \textsc{Union} operations are given in Alg.~\ref{alg:disjoint-forest}.
\vspace{-2ex}
\section{Literature and Misconceptions}\label{sec:rel}

\noindent In this section, we present a detailed review of related work on peeling algorithms.
We point some misconceptions about the definitions and the consequences.
Our focus is on peeling algorithms and their output, so we limit our scope to $k$-core and $k$-truss decompositions and their generalizations.
Detailed literature review of dense subgraph discovery can be found in~\cite{Gionis15,Lee10}.

\subsection{k-core decomposition}

\noindent The very first definition of a $k$-core related concept is given by Erd\H{o}s and Hajnal~\cite{ErHa66} in 1966.
They defined the degeneracy as the largest maximum core number of a vertex in the graph.
Matula introduced the min-max theorem~\cite{Matula68} for the same thing, highlighting the relationship between degree orderings of the graph and the minimum degree of any subgraph, and its applications to graph coloring problem.
Degeneracy number has been rediscovered numerous times in the context of graph orientations and is alternately called the coloring number~\cite{LiWh70}, and linkage~\cite{Freuder82}.

First definition of the $k$-core subgraph is given by Seidman~\cite{Seidman83} for social networks analysis, and also by Matula and Beck~\cite{MaBe83}, as $k$-linkage, for clustering and graph coloring applications, in the same year of 1983.
Seidman~\cite{Seidman83} introduced the core collapse sequence, also known as degeneracy ordering of vertices, as an important graph feature.
He states that $k$-cores are good \emph{seedbeds} that can be used to find further dense substructures.
Though, there is no algorithm in~\cite{Seidman83} on how to find the $k$-cores.
Matula and Beck~\cite{MaBe83}, on the other hand, gives algorithms for finding $\kval_2$ values of vertices, and also finding all the $k$-cores of a graph (and their hierarchy) by using these $\kval_2$ values, because there can be multiple $k$-cores for same $k$ value.
Both papers defined the $k$-core subgraph as follows:
\vspace{-1ex}
\begin{quote}
\emph{``A \textbf{connected} and maximal subgraph $H$ is $k$-core ($k$-linkage) if every vertex in $H$ has at least degree $k$.''}~\cite{Seidman83, MaBe83}
\end{quote}

\begin{figure}[!h]
\vspace*{1ex}
\centering
 \includegraphics[width=0.69\linewidth]{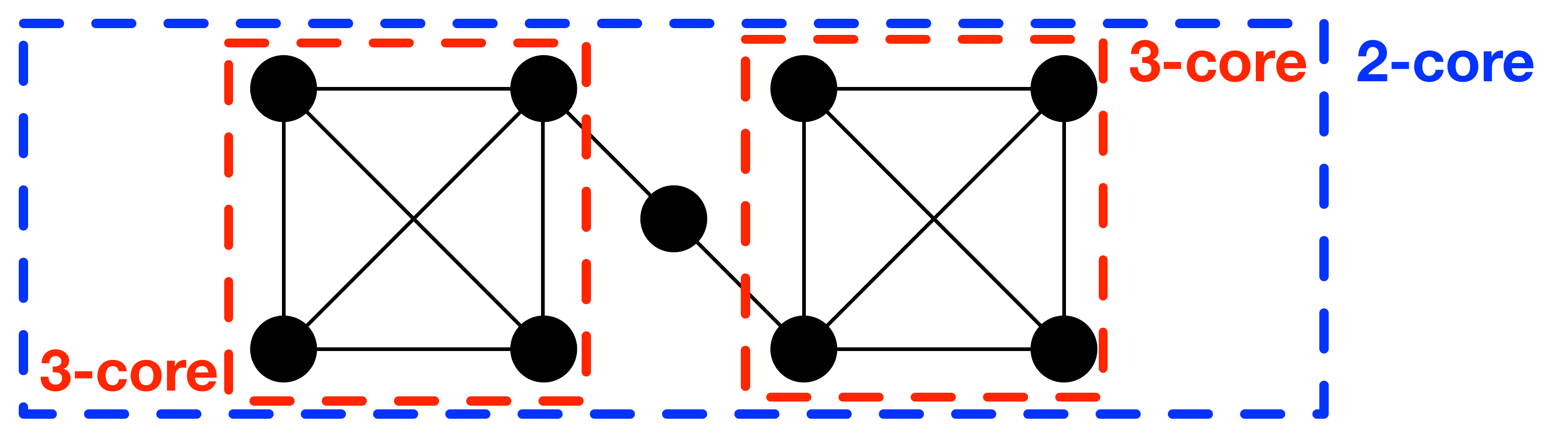}
 \vspace*{-2.5ex}
\linespread{0}\selectfont
\caption{\small Multiple $3$-cores}
\linespread{1}\selectfont
 \vspace*{1ex}
\label{fig:conn} 
\end{figure}

\noindent The \emph{connectedness} is an important detail in this definition because it requires a post-processing traversal operation on vertices to locate all the $k$-cores of the graph.
Figure~\ref{fig:conn} shows this.
There are two 3-cores in the graph, and there is no way to distinguish them at the end of the peeling process by just looking at the $\kval$ values of vertices.

Batagelj and Zaversnik introduced an efficient implementation that uses bucket data structure to find the $\kval$ values of vertices~\cite{BaZa03}.
They defined the $k$-core as a not necessarily connected subgraph, in contrast to previous work they cited~\cite{Seidman83, MaBe83}.
With this assumption, they claimed that their implementation finds all the $k$-cores of the graph.

Finding the relationships between $k$-cores of a graph has gained a lot of interest.
Nested structure of $k$-cores reveals a hierarchy, and it has been shown to be  useful for visualization~\cite{Alvarez06} and understanding the underlying structure of complex networks arising in many domains.
Carmi \emph{et al.}~\cite{Carmi07} and Alvarez-Hamelin \emph{et al.}~\cite{Alvarez08} investigated the $k$-core hierarchy of internet topology at autonomous systems (AS) level.
Healy \emph{et al.}~\cite{Healy08} compared the $k$-core hierarchies of real-world graphs in different domains and some generative models.

Given the practical benefit and efficiency of $k$-core decomposition, there has been a lot of recent work to adapt $k$-core algorithms for different data types or setups.
Out of memory computation is an important topic for many graph analytic problems that deal with massive graphs not fitting in memory.
Cheng \emph{et al.}~\cite{Cheng11} introduced the first external-memory algorithm.
Wen \emph{et al.}~\cite{Wen16} and Khaouid \emph{et al.}~\cite{Khaouid15} provided further improvements in this direction.
Regarding the different type of graphs, Giatsidis \emph{et al.} adapted the $k$-core decomposition for weighted~\cite{Giatsidis13} and directed~\cite{Giatsidis11} graphs.
To handle the dynamic nature of real-world data, Sariyuce \emph{et al.}~\cite{Sariyuce13-VLDB} introduced the first streaming algorithms to maintain $k$-core decomposition of graphs upon edge insertions and removals.
They recently improved these algorithms further by leveraging the information beyond 2-hop~\cite{Sariyuce16}.
Li \emph{et al.}~\cite{Li14} also proposed incremental algorithms for the same problem.
More recently, Wu \emph{et al.}~\cite{Wu15} approached dynamic data from a different angle, and adapted $k$-cores for temporal graphs where possibly multiple interactions between entities occur at different times.
Motivated by the incomplete and uncertain nature of the real network data, O'Brien and Sullivan~\cite{OBrien14} proposed new methods to locally estimate core numbers ($\kval$ values) of vertices when entire graph is not known, and Bonchi \emph{et al.}~\cite{Bonchi14} showed how to efficiently do the $k$-core decomposition on uncertain graphs, which has existence probabilities on the edges.

\begin{figure}[!]
\vspace*{-1ex}
\centering
\linespread{0}\selectfont
\caption{\small $k$-dense~\protect\cite{Saito06} (triangle $k$-core~\protect\cite{Zhang12}), $k$-truss~\protect\cite{Cohen08} ($k$-community~\protect\cite{Verma12}) and $k$-truss community~\protect\cite{Huang14} ($k$-$(2,3)$ nucleus~\protect\cite{Sariyuce15}) on the same graph for $k$=2. Each subgraph given by the corresponding algorithm is shown in dashed.}
\linespread{1}\selectfont
\vspace{0.5ex}
\includegraphics[width=0.69\linewidth]{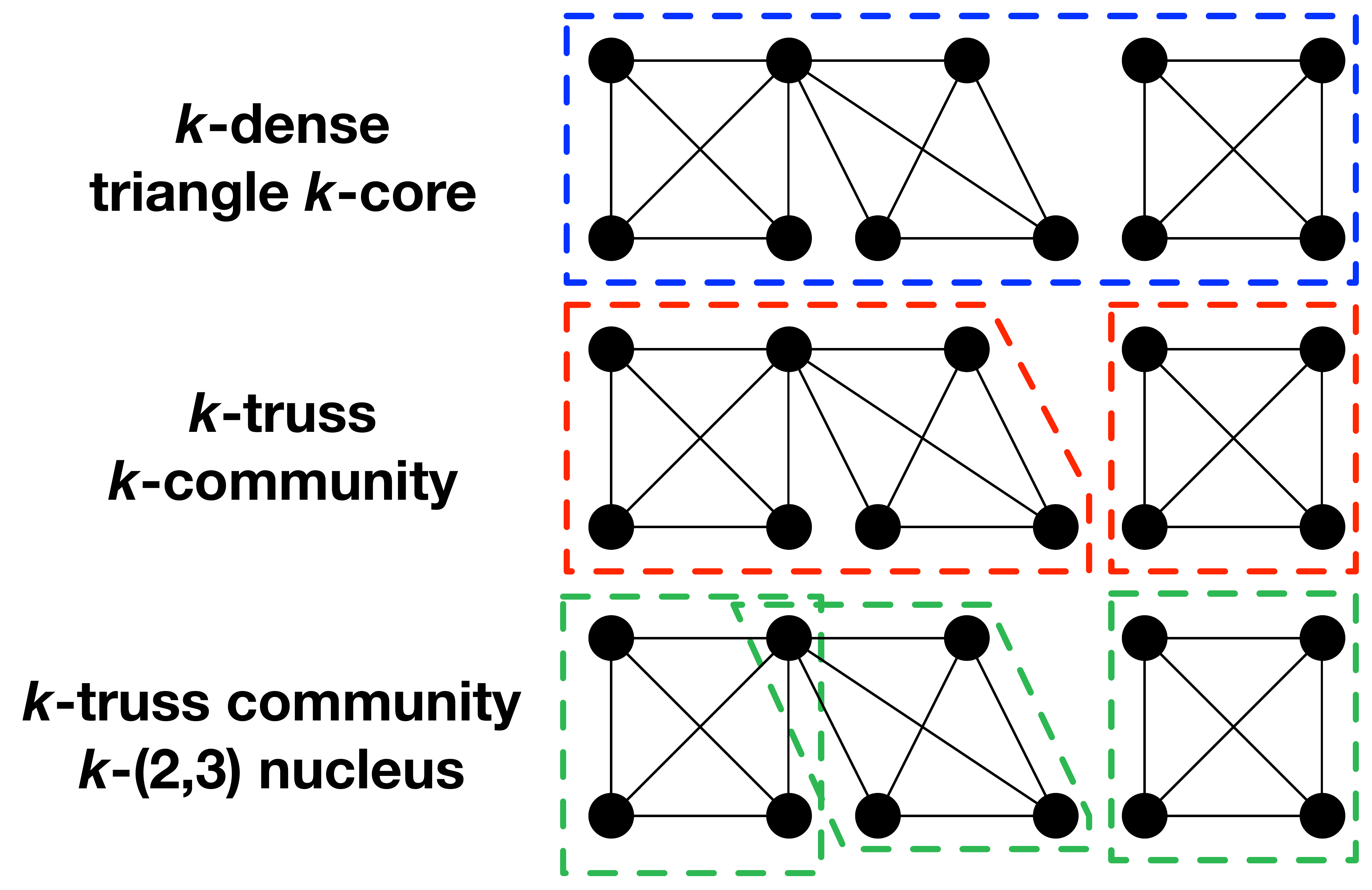}
\label{fig:ktruss}
\end{figure}

One common oversight in all those recent work (except ~\cite{Sariyuce13-VLDB,Sariyuce15}) is that they ignore the \textbf{connectivity} of $k$-cores.
This oversight does not change their results, but limit their contributions: they adapt/improve the peeling part of $k$-core decomposition, which finds the $\kval$s of vertices, not the entire $k$-core decomposition which also needs traversal to locate all the (connected) $k$-cores.
Considering the external memory $k$-core decomposition algorithms~\cite{Cheng11,Khaouid15,Wen16}, existing works only focused on how to compute the $\lambda$ values of vertices. However, the additional traversal operation in external memory is not taken into consideration which is at least as expensive as finding $\lambda$ values. Finding the (connected) $k$-cores and constructing the hierarchy among them efficiently in the external memory computation model is not a trivial problem and will limit the performance of proposed algorithms for finding $k$-core subgraphs and constructing the hierarchy. Similar argument can be considered for weighted~\cite{Giatsidis13}, probabilistic~\cite{Bonchi14}, and temporal~\cite{Wu15} $k$-core decompositions, all of which have some kind of threshold-based adaptations on weights, probabilities and timestamps, respectively. On the other hand, connectedness definition is semantically unclear for some existing works like the directed graph core decomposition~\cite{Giatsidis11}. It is only defined that in- and out-degrees of vertices can be considered to find two $\lambda$ values, but traversal semantic is not defined for finding subgraphs or constructing the hierarchy. One can think about building the hierarchy by considering the edges from lower level $k$-cores to higher level ones, or the opposite.
To remedy those misconceptions, we focus on the efficient computation of traversal part for $k$-core decomposition and its higher-order variants.

\subsection{k-truss decomposition}

\noindent $k$-truss decomposition is inspired by the $k$-core and can be thought as the same peeling problem in a higher level that deals with triangles.
It is independently introduced, with subtle differences, by several researchers.
Chronologically, the idea is first proposed by Saito \emph{et al.}~\cite{Saito06}, to the best of our knowledge, in 2006:
\begin{quote}
\emph{``$\textbf{k}$\textbf{-dense} is a subgraph $S$ if each adjacent vertex pair in $S$ has more than or equal to ($k$-2) common adjacent vertices in $S$.''}
\end{quote}
In other words, each edge in $S$ should be involved in at least $k$-2 triangles.
Nothing is mentioned about the connectedness of the vertices and edges, which implies that a $k$-dense subgraph might have multiple components.
Saito \emph{et al.} argue that $k$-dense is a good compromise between easy to compute $k$-cores and high quality $k$-cliques, and it is useful to detect communities in social networks.
In 2008, Jonathan Cohen introduced the \emph{$k$-truss} as a better model for cohesive subgraphs in social networks~\cite{Cohen08}, which became the most popular naming in the literature:
\begin{quote}
\emph{``$\textbf{k}$\textbf{-truss} is a one-component subgraph such that each edge is reinforced by at least $k$-2 pairs of edges making a triangle with that edge.''}
\end{quote}
In 2012, Zhang and Parthasarathy~\cite{Zhang12} proposed a new definiton for visualization purposes:
\begin{quote}
\emph{``\textbf{triangle }$\textbf{k}$\textbf{-core} is a subgraph that each edge is contained within at least $k$ triangles in the subgraph.''}
\end{quote}
Again there was no reference to the connectedness, implying multiple components can be observed in a triangle $k$-core.
In the same year, Verma and Butenko~\cite{Verma12} introduced the following:

\begin{quote}
\emph{``$\textbf{k}$\textbf{-community} is a connected subgraph if every edge is involved in at least $k$ triangles.''}
\end{quote}

The subtle difference between those papers is the connectedness issue.
$k$-dense~\cite{Saito06} and triangle $k$-core~\cite{Zhang12} definitions allow the subgraph to be disconnected whereas the $k$-truss~\cite{Cohen08} and $k$-community~\cite{Verma12} are defined to be connected.
All of these works only provided algorithms to find the $\kval_3$ values of edges.
$k$-dense and triangle $k$-cores can be found this way since they can be disconnected.
However, finding the $k$-truss and $k$-community subgraphs requires a post-processing traversal operation, which increases the cost.
As a stronger alternative to the $k$-truss, Huang \emph{et al.}~\cite{Huang14} introduced the \emph{$k$-truss community}.
The only difference is that each edge pair in a $k$-truss community is directly or transitively triangle-connected, where two edges should reside in the same triangle to be triangle-connected.
The generic $k$-$(r,s)$ nucleus, proposed by Sariyuce \emph{et al.}~\cite{Sariyuce15},  for $r=2, s=3$ gives the exact same definition.
This brings a stronger condition on the connectivity structure, and shown to result in denser subgraphs than the classical $k$-truss definition~\cite{Huang14}.
However, it has an extra overhead of post-processing traversal operation that requires to visit triangles, which is more expensive than the traditional traversal.
Authors devised TCP index, a tree structure at each vertex, to remedy this issue~\cite{Huang14}.
Figure~\ref{fig:ktruss} highlights the difference between those definitions on a simple example.

$k$-truss decomposition serves as a better alternative to the $k$-core.
For most applications that $k$-core is useful for, $k$-truss decomposition performs better.
Gregori \emph{et al.}~\cite{Gregori11} investigated the structure of internet AS-level topologies by looking at the $k$-dense subgraphs, similar to Carmi \emph{et al.}~\cite{Carmi07} and Alvarez-Hamelin \emph{et al.}~\cite{Alvarez08} who used $k$-core for same purpose.
Orsini \emph{et al.}~\cite{Orsini14} also investigated the evolution of $k$-dense subgraphs in AS-level topologies.
It has been also used to understand the global organization of clusters in complex networks~\cite{Colomer13}.
Colomer-de-Simon \emph{et al.} used the hierarchy of $k$-dense subgraphs to visualize real-world networks, as Healy \emph{et al.}~\cite{Healy08} used the $k$-cores for the same objective.

Proven strength of $k$-truss decomposition drew further interest for adapting to different data types and setups, similar to the $k$-core literature.
Wang and Cheng introduced external memory algorithms~\cite{WaCh12} and more improvements are provided by Zhao and Tung~\cite{Zhao13} for visualization purposes.
More recently, Huang \emph{et al.}~\cite{Huang16} introduced probabilistic truss decomposition for uncertain graphs.

Similar to the $k$-core case, overlooking the connectivity constraints limits the contributions in the $k$-truss literature as well. For example, external memory $k$-truss decomposition~\cite{WaCh12} would be more expensive and require more intricate algorithms if it is done to find connected subgraphs by doing the traversal in external memory model. We believe that our algorithms for efficiently finding the $k$-trusses and constructing the hierarchy will be helpful to deal with this issue.

\subsection{Generalizations}\label{subsec:gener}
\noindent Given the similarity between $k$-core and $k$-truss decompositions, people have been interested in unified schemes to generalize the peeling process for a broader set of graph substructures.

Saito \emph{et al.} pointed a possible direction of generalization in their pioneering work~\cite{Saito06}, where they defined $k$-dense subgraphs.
Their proposal is as follows:
\begin{quote}
\emph{``Subgraph $S$ is a $\textbf{h}$\textbf{-level} $\textbf{k}$\textbf{-dense community} if the vertices in every $h$-clique of $S$ is adjacent to at least $h$-$k$ common vertices.''}~\cite{Saito06}
\end{quote}
In other words, $h$-level $k$-dense community is the set of $h$-cliques where each $h$-clique is contained in at least $h-k$ number of $(h+1)$-cliques.
Note that, there is no connectivity constraint in the definition.
$h$-level $k$-dense community subsumes the disconnected $k$-core, which contains multiple $k$-cores, for $h=1$.
For $h=2$, it is their $k$-dense definition~\cite{Saito06}.
They claimed that $h$-level $k$-dense communities for $h>2$ are more or less same with $h=2$ and incurs higher computation cost.
So they did not dive into more algorithmic details and stick with $h=2$.

Sariyuce \emph{et al.}~\cite{Sariyuce15} introduced a broader definition to unify the existing proposals, which can be found by a generic peeling algorithm.
As explained in Section~\ref{subsec:nucleus}, their definition subsumes $k$-core and $k$-truss community~\cite{Huang14} concepts.
It is also more generic than $h$-level $k$-dense community of~\cite{Saito06}, since (1) it allows to look for involvement of cliques whose size can differ by more than one, (2) enforces a stronger connectivity constraint to get denser subgraphs.
$h$-level $k$-dense community can be expressed as the $k$-$(r,r+1)$ nucleus which does not have any connectivity constraint ($k$ is actually $h-k$ and it does not matter).
Well-defined theoretical notion of $k$-$(r,s)$ nucleus enables to provide a unified algorithm to find all the nuclei in graph, as explained in Section~\ref{subsec:nucleus}.

Sariyuce \emph{et al.}~\cite{Sariyuce15} also analyzed the time and space complexity of $(r,s)$-nucleus decomposition.
For the first phase, they report that finding $\kval$ values of nuclei (Alg.~\ref{alg:set-nucleus}) requires $O(RT_r(G)+\sum_v \omega_r(v)d(v)^{s-r})$ time with $O(|K_r(G)|)$ space, where $RT_r(G)$ is  $K_r$ enumeration time, and second part is searching each $K_s$ that a $K_r$ is involved in ($\omega_r(v)$ is the number of $K_r$s containing vertex $v$, $d(v)$ is the degree of $v$, and $|K_r(G)|$ is the number of $K_r$s in $G$).
For the second phase, traversal on the entire graph needs to access each $K_r$ and examine all the $K_s$s it is involved.
Its time complexity is the same as the second part of first phase: $O(\sum_v \omega_r(v)d(v)^{s-r})$ which also gives the total time complexity.
\vspace{-2ex}

\section{Algorithms}\label{sec:alg}
\noindent In this part, we first highlight the challenging aspects of the traversal phase, then introduce two algorithms for faster computation of $(r,s)$-nucleus decomposition to meet those challenges. 

\subsection{Challenges of traversal}\label{subsec:cha}

\begin{figure}[!b]
\centering
\linespread{0.1}\selectfont{}
\includegraphics[width=0.99\linewidth]{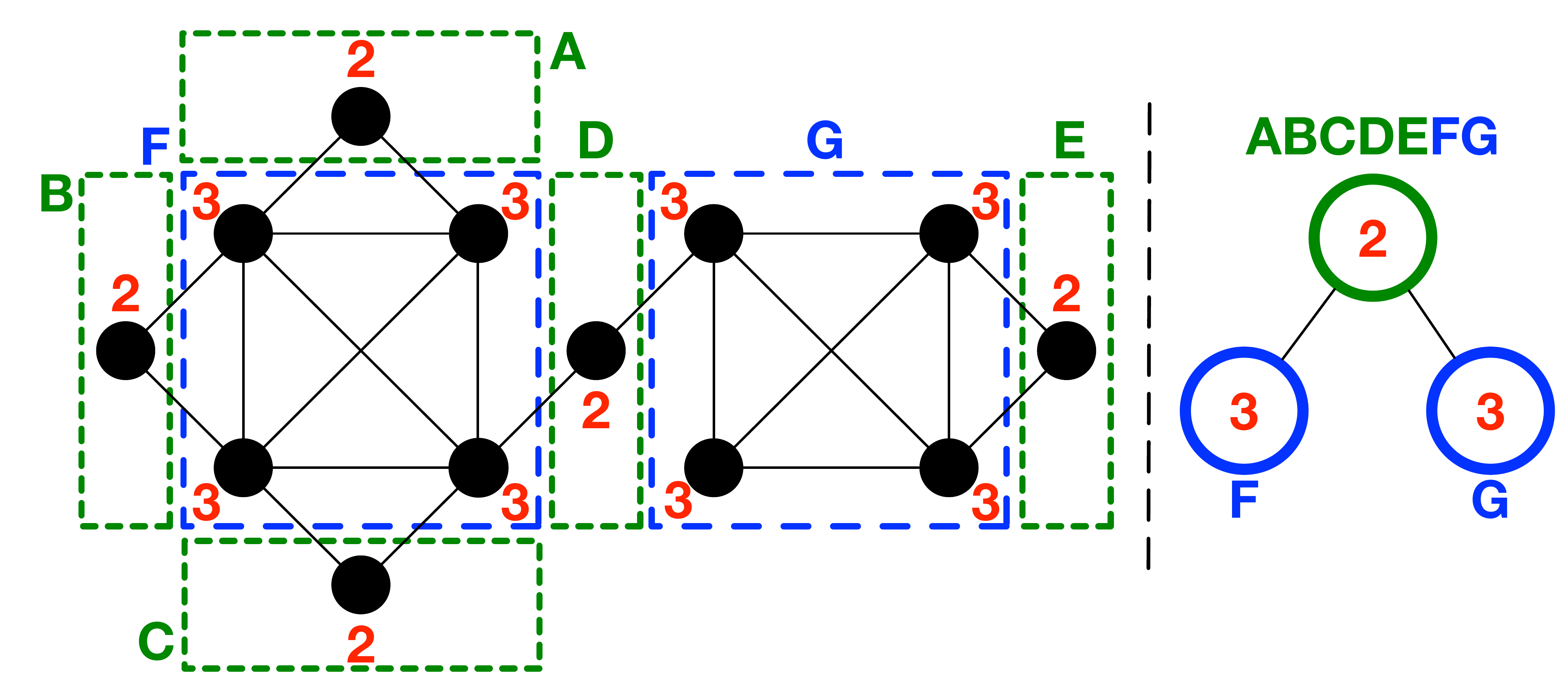}
 \vspace*{-2.0ex}
\linespread{0}\selectfont
\caption{\small Example of $T_{1,2}$s for $\kval=2$ and $\kval=3$. Hierarchy tree is shown on the right with participating $T_{1,2}$s. $\kval$s are shown in red. Traversal algorithm needs to infer that, for instance, components A and E are in the same 2-core.}
\linespread{1}\selectfont
\label{fig:challenge}
 \vspace*{-1ex}
\end{figure}

\noindent As mentioned in the previous section, time complexity of the traversal algorithm for $(r,s)$ nuclei is $O(\sum_v \omega_r(v)d(v)^{s-r})$.
However, designing an algorithm that constructs the hierarchy with this complexity is challenging.
In~\cite{Sariyuce15}, it is stated that finding the nuclei in the reverse order of $\kval$ is better since it enables to discover previously found components, thus avoiding repetition.
No further details are given, though.
This actually corresponds to finding all $T_{r,s}$ (sub-$(r,s)$ nuclei of Definition~\ref{def:sub-nucleus}), connected $K_r$s with the same $\kval$ value, and incorporating the relations among them.
But, keeping track of all the $T_{r,s}$ in a structured way is hard.
Figure~\ref{fig:challenge} shows a case for $k$-core ($r=1,s=2$).
Traversal algorithm needs to understand the relation between $T_{1,2}$s of equal $\kval$ that are not directly connected.
For instance, $T_{r,s}$ A and E are in the same 2-core, but the traditional BFS will find 3 other $T_{r,s}$ (F, D, G) between those two.
During the traversal operation, there is a need to detect each $k$-$(r,s)$ nucleus, determine containment relations and construct the hierarchy.
One solution that can be thought is to construct the - expectedly smaller - supergraph which takes all the $T_{r,s}$ as vertices and their connections as edges.
Then, repetitive traversals can be performed on this supergraph to find each $k$-$(r,s)$ nucleus and the hierarchy.
However, it is not guaranteed to get a significantly smaller supergraph which can be leveraged for repetitive traversal.
The $T_{r,s}$ structure of real-world networks, which are investigated in Section~\ref{sec:exp}, also verify this concern.
It is clear that there is a need for a lightweight algorithm/data structure that can be used on-the-fly, so that all the $k$-$(r,s)$ nuclei can be discovered with the hierarchy during the traversal algorithm.

The other challenge with the traversal algorithm is the high computational cost for $r \ge 2$ cases.
Consider the $(2,3)$ case.
We need to traverse on edges, and determine the adjacencies of edges by looking at their common triangles.
At each step, it requires to find the common neighbors of the vertices on extremities (of the edge), check whether each adjacent edge is previously visited, and push to queue if not.
As explained at the end of Section~\ref{subsec:gener}, complexity becomes $O(3*|\triangle|)$.
Cost is getting much higher if we look for $(3,4)$ nuclei, which is shown to give denser subgraphs with more detailed hierarchy.
Ideally we are looking to completely avoid the costly traversal operation.

\subsection{Disjoint-set forest idea}\label{subsec:dstraversal}

\noindent We propose to use disjoint-set forest data structure (DF) to track the disjoint $T_{r,s}$ (of equal $\kval$), and construct the hierarchy where $T_{r,s}$ with smaller $\kval$ is on the upper side, and greater $\kval$ is on the lower side.
DF has been used to track connected components of a graph and fits perfectly to our problem where we need to find the connected components at multiple levels.

\begin{algorithm}[!b]
\small
\caption{\textsc{DF-Traversal($G, r, s, \kval(\cdot), max\kval$)}}
\label{alg:DFtraverse-nucleus}  
  $\ahrc \leftarrow$ list of \tsns~in a tree structure\; \label{ln:df1}
  $\comp{.}~~\forall~K_r \in G$ \tcp*{\tsn~id for each $K_r$} \label{ln:df2}  
  $\visited{v} = \mfalse$ \tcp*{lazy init}\label{ln:df3}
  \For{\meach $k \in [max\kval, 1]$}{\label{ln:df4}
    \For{\meach $K_r~u \in G$}{\label{ln:df5}
      \If{$\kval(u) = k \mand \mnot\visited{u}$}{\label{ln:df6}
        \textsc{SubNucleus} ($u, G, r, s, \kval(\cdot), {\tt visited}, {\tt comp}, \ahrc$)\label{ln:df7}
      }
    }
  }
  $root \leftarrow$ \tsn, with $\kval=0$\;\label{ln:df8}
  \For{\meach $s \in \ahrc$}{\label{ln:df9}
    \lIf{$s$.\textnormal{parent} is null}{$s$.parent $\leftarrow root$}\label{ln:df10}
  }
  $\ahrc.\tadd$($root$) and \textsc{Report All the Nuclei by \ahrc,\tcomp}\; \label{ln:df11}
\end{algorithm}

\begin{algorithm}[!b]
\small
\caption{\textsc{SubNucleus($u, G, r, s, \kval(\cdot), {\tt visited}, {\tt comp}, \ahrc$)}} 
\label{alg:subnucleus}  
  $sn \leftarrow$ \tsn, with $\kval=\kval(u)$, $\ahrc.\tadd$($sn$)\; \label{ln:sn1}
  $\comp{u} \leftarrow sn$, $k \leftarrow \kval(u)$ \; \label{ln:sn2}
  $\marked{v} = \mfalse$ \tcp*{lazy init} \label{ln:sn3}
  $merge \leftarrow$ list of \tsns, $merge.\tadd(sn$)\; \label{ln:sn4}
  $Q \leftarrow$ empty queue, $Q$.push($u$)\; \label{ln:sn5}
  $\visited{u} \leftarrow \mtrue$ \; \label{ln:sn6}
  \While{$\mnot Q$\textnormal{.empty()}}{ \label{ln:sn7}
    $u \leftarrow Q$.pop(), $\comp{u} \leftarrow sn$\; \label{ln:sn8}
    \For{\meach $K_r$ $v$ s.t.\\ \label{ln:sn9}
    \nonl $(u\cup v\subset C) \land (C$ is a $K_s) \land (\kval_{r,s}(C) = k)$}{
      \If{$\kval(v) = k$}{ \label{ln:sn10}
        \If{$\mnot\visited{v}$}{ \label{ln:sn11}          
          $Q$.push($v$), $\visited{v} \leftarrow \mtrue$ \; \label{ln:sn12}
	  $\comp{v} \leftarrow sn$\; \label{ln:sn13}        
        } 
      }
      \Else { \label{ln:sn14}
        $s \leftarrow \comp{v}$ \tcp{$\kval(v) > k$} \label{ln:sn15}
        \If{$\mnot\marked{s}$}{ \label{ln:sn16}
          $\marked{s} = \mtrue$\; \label{ln:sn17}
          $s \leftarrow \textsc{Find-r} (s)$\; \label{ln:sn18}
          \If{$\mnot\marked{s} \land s \neq sn$}{ \label{ln:sn19}
            \If{$\hrc{s}.\kval > k$}{ \label{ln:sn20}
              $\hrc{s}$.parent $ \leftarrow \hrc{s}$.root $\leftarrow sn$ \label{ln:sn21}
            }
            \lElse {$merge.\tadd(s)$~~~~~~~~~~{{\tt// $\hrc{s}.\kval = k$}}} \label{ln:sn22}
            $\marked{s} = \mtrue$\; \label{ln:sn23}
          }   
        }
      }
    }
  }
  \For{\meach $m,n \in$ merge} { \label{ln:sn24}
    \textsc{Union-r}($m,n$) \label{ln:sn25}
  }
\end{algorithm}

\begin{figure}[!t]
\centering
\vspace*{-2ex}
\linespread{0}\selectfont{}
\caption{\small A graph shown with $T_{r,s}$ regions on the left and the corresponding hierarchy-skeleton on the right. $\kval$ values of  $T_{r,s}$s are the white numbers. Thin edges are disjoint-set forests.}
\linespread{1}\selectfont
 \includegraphics[width=0.99\linewidth]{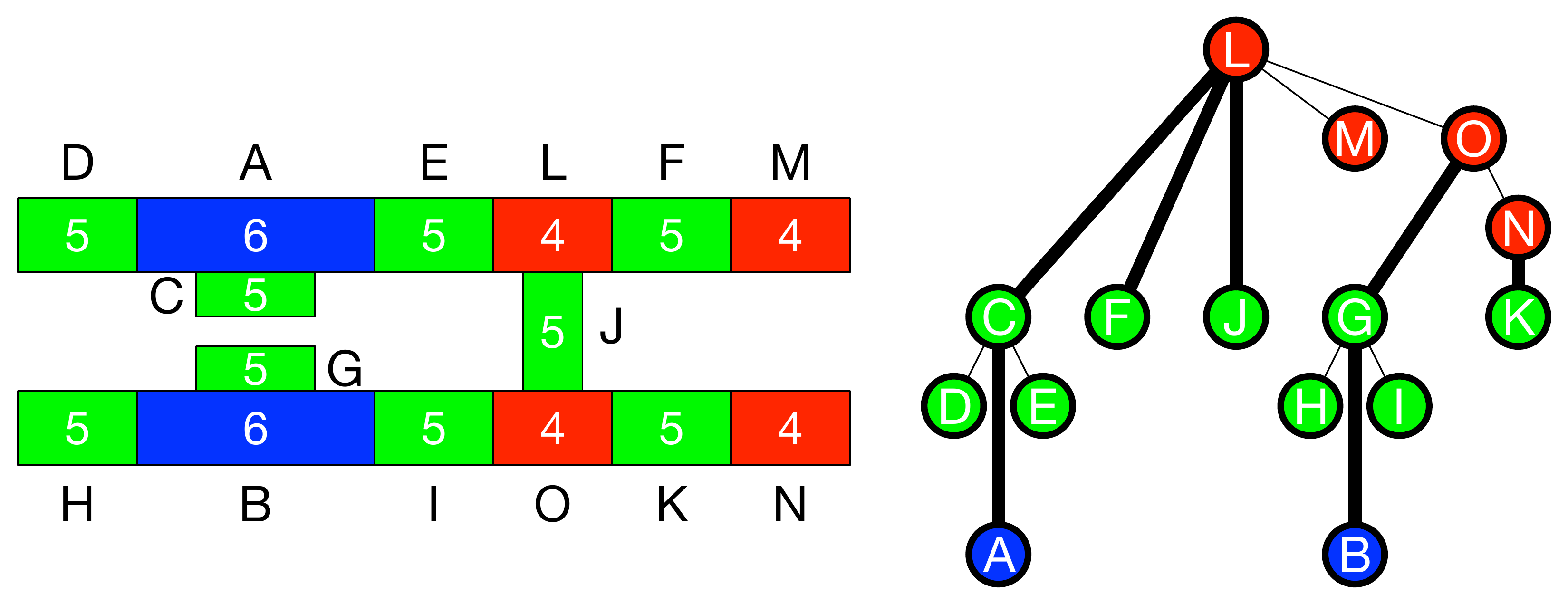}
\label{fig:hier}
\vspace*{-2ex}
\end{figure}

\textsc{DF-Traversal} algorithm, outlined in Alg.~\ref{alg:DFtraverse-nucleus}, is used to replace the naive \textsc{Traversal} (Alg.~\ref{alg:traverse-nucleus}) in \textsc{NucleusDecomposition} (Alg.~\ref{alg:naive}).
Basically it finds all the $T_{r,s}$ in the decreasing order of $\kval$.
We construct the \emph{hierarchy-skeleton} tree by using $T_{r,s}$s.
Each node in the hierarchy-skeleton is a $T_{r,s}$.
We define \tsn~struct to represent a $T_{r,s}$.
It consists of $\kval$, {\tt rank, parent} and {\tt root} fields.
$\kval$ field is the $\kval(v)$ for $v \in$ $T_{r,s}$, \trank is the height of the node in the hierarchy-skeleton, \tparent is a pointer to the parent node and \troot is a pointer to the root node of the hierarchy-skeleton.
Default values for \tparent and \troot are \tnull, and \trank is 0.
Figure~\ref{fig:hier} shows an example hierarchy-skeleton obtained by Alg.~\ref{alg:DFtraverse-nucleus}.
Thin edges show the disjoint-set forests consisting of  $T_{r,s}$s of equal $\kval$ value.
The hierarchy of all $(r,s)$ nuclei, the output we are interested, can be obtained by using the hierarchy-skeleton easily: we just take the child-parent links for which the $\kval$ values are different.

In the \textsc{DF-Traversal} algorithm, we keep all the \tsns\\
in \thrc list (line~\ref{ln:df1}) which also represents the hierarchy-skeleton.
$K_r$s in a  $T_{r,s}$ are stored by inverse-indices; \tcomp keeps\\ \tsn~index of each $K_r$ in \thrc (line~\ref{ln:df2}).
We also use \tvisited to keep track of traversed $K_r$s (line~\ref{ln:df3}).
Main idea is to find each $T_{r,s}$ in decreasing order of $\kval$ (in lines~\ref{ln:df4}-\ref{ln:df7}).
We construct the hierarchy-skeleton in a bottom-up manner this way and it lets us to use DF to find the representative $T_{r,s}$, i.e., the greatest ancestor, at any time.
At each iteration we find an un-\tvisited $K_r$ with the $\kval$ value in order (line~\ref{ln:df6}) and find its $T_{r,s}$ by \textsc{SubNucleus} algorithm (line~\ref{ln:df7}), which also updates the hierarchy-skeleton.

\textsc{SubNucleus} (Alg.~\ref{alg:subnucleus}) starts by creating a \tsn, with $\kval$ of the $K_r$ of interest.
We will store the discovered $K_r$s in this \tsn~(by inverse indices).
We put this \tsn~into \thrc (line~\ref{ln:sn1}) and assign its \tcomp id (line~\ref{ln:sn2}).
We use \tmarked (line~\ref{ln:sn3}) to mark the adjacent \tsns~encountered during traversal so that unnecessary computation is avoided in lines~\ref{ln:sn16}-\ref{ln:sn23}.
We do traversal (lines~\ref{ln:sn7}-\ref{ln:sn23}) by using a queue.
At each step of the traversal, we process the next $K_r$ in the queue.
First, we assign its \tcomp id as the new \tsn~(line~\ref{ln:sn8}) and then visit the adjacent $K_r$s residing in same $K_s$s, in which min $\kval$ of $K_r$ is equal to  $\kval(v)$ (line~\ref{ln:sn9}).
This is exactly the condition for $T_{r,s}$, given in Definition~\ref{def:sub-nucleus}.
For each adjacent $K_r$, its $\kval$ is either equal (line~\ref{ln:sn10}) or greater (line~\ref{ln:sn14}) by definition of $T_{r,s}$.
If it is equal and not \tvisited before, we visit it, put into queue (line~\ref{ln:sn12}) and also store in the current \tsn~(line~\ref{ln:sn13}).
Otherwise, we find an adjacent \tsn~$s$ with greater $\kval$ (line~\ref{ln:sn15}), that is already in hierarchy-skeleton, and can update the hierarchy-skeleton (lines~\ref{ln:sn18}-\ref{ln:sn23}) unless we had encountered $s$ before (line~\ref{ln:sn16}).

Location of the \tsn~$s$ in the hierarchy-skeleton is important.
If it is parentless, we can just make it a child of the current \tsn~ we build.
If not, it means \tsn~$s$ is a part of a larger structure and we should relate our current \tsn~to the representative of this large structure, which is the greatest ancestor of $s$ that is guaranteed to have greater or equal $\kval$ (by line~\ref{ln:sn9}).
So, hierarchy-skeleton update starts by finding the greatest ancestor of the $s$ in line~\ref{ln:sn18}.
\textsc{Find-r} procedure is defined in Alg.~\ref{alg:newdisjoint-forest}.
Its difference from \textsc{Find} of Alg.~\ref{alg:disjoint-forest} is that we use \troot field, not {\tt parent}.
\troot of a node implies its greatest ancestor in the hierarchy-skeleton, i.e., either it is the greatest ancestor or a few steps of \troot iterations would find the greatest ancestor.
\tparent of a node, on the other hand, represents the links in hierarchy-skeleton, and not modified in \textsc{Find-r}.
After finding the root and making sure that it is not processed before (line~\ref{ln:sn19}), we can merge the current \tsn~to the hierarchy-skeleton.
If the root has greater $\kval$, we make it a child of our current \tsn~(line~\ref{ln:sn21}), by assigning both \troot and \tparent fields.
Otherwise, we defer merging to the end (line~\ref{ln:sn22}), where we merge all \tsns~with equal $\kval$ by \textsc{Union-r} operations (lines~\ref{ln:sn24}-\ref{ln:sn25}), defined in Alg.~\ref{alg:newdisjoint-forest}.
\textsc{Union-r} is slightly different than \textsc{Union} of Alg.~\ref{alg:disjoint-forest} in that it uses \textsc{Find-r} instead of \textsc{Find} and sets the \troot field of child node to the parent (in \textsc{Link-r}).

Figure~\ref{fig:hier} displays the resulting hierarchy-skeleton for the $T_{r,s}$ regions shown on the left.
We process $T_{r,s}$ in alphabetical order, which also conforms with decreasing order of $\kval$.
Consider the $T_{r,s}$ O, which is found and processed last.
O finds the adjacent $T_{r,s}$s I, J and K, during lines~\ref{ln:sn9}-\ref{ln:sn23} of Alg.~\ref{alg:subnucleus}.
All have greater $\kval$ values, so we focus on lines~\ref{ln:sn14}-\ref{ln:sn23}.
Greatest ancestor of I is G, and we make G child of O (line~\ref{ln:sn21}) since its $\kval$ is greater.
Greatest ancestors of J and K are L and N, respectively, and they have equal $\kval$ values.
So, we merge L and N with O in lines~\ref{ln:sn24}-\ref{ln:sn25}.
Say we merge O and N first and O becomes \tparent of N since its rank is higher.
Then, we merge O and L.
Their ranks are equal and we arbitrarily choose L as the {\tt parent}.

After traversing all the $T_{r,s}$, we create a root \tsn~to represent entire graph and make it parent to all the parentless nodes (lines~\ref{ln:df8}-\ref{ln:df11} in Alg.~\ref{alg:DFtraverse-nucleus}).
Time complexity of \textsc{DF-Traversal} does not change, i.e., $O(\sum_v \omega_r(v)d(v)^{s-r})$.
Additional space is required by the auxiliary data structures in \textsc{DF-Traversal}.
\ahrc~needs $4\cdot|T_{r,s}|$ (for four fields), and \tcomp and {\tt visited} requires $|K_r|$ space each.
In addition, \textsc{SubNucleus} might require at most $2\cdot|T_{r,s}|$ for {\tt marked} and \textit{merge}, and at most $|K_r|$ for \textit{Q}, but reaching those upper bounds in practice is quite unlikely.
Overall, additional space requirement of \textsc{DF-Traversal} at any instant is between $4\cdot|T_{r,s}|+2\cdot|K_r|$ and $6\cdot|T_{r,s}|+3\cdot|K_r|$.
An upper bound for $|T_{r,s}|$ can be given as $|K_r|$, when each $K_r$ is assumed to a \tsn, but this case is also quite unlikely as we show in Section~\ref{sec:exp}.

\subsection{Avoiding traversal}\label{subsec:ditchtraversal}

\LinesNumberedHidden
\begin{algorithm}[!b]
\small
\caption{\textsc{New Disjoint-Set Forest}}
\label{alg:newdisjoint-forest}
\underline{\textsc{Link-r}}($x, y$): \tcp*{$x$ and $y$ are nodes in the tree}
~\lIf{$x$.\tn{rank} $> y$.\tn{rank}}{$y$.{\tt parent} $\leftarrow x$, $y$.{\tt root} $\leftarrow x$}
~\Else {
~  $x$.{\tt parent} $\leftarrow y$, $x$.{\tt root} $\leftarrow y$\;
~  \lIf{$x$.\tn{rank} $= y$.\tn{rank}}{$y$.\tn{rank} $\leftarrow y$.\tn{rank} + 1}}
\underline{\textsc{Find-r}}($x$):\;
~$S \leftarrow$ empty set\;
~\lWhile{$x.{\tt root}~is~not~null$}{$x \leftarrow x.{\tt root}$,  $S$.add($x$)}        
~\lFor{\meach $u \in S$}{$u.{\tt root} \leftarrow x$}
~\Return x\;
\underline{\textsc{Union-r}}($x, y$):~\textsc{Link-r(Find-r($x$), Find-r($y$))}
\end{algorithm}
\LinesNumbered

\begin{algorithm}[!b]
\small
\caption{\textsc{FastNucleusDecomposition($G, r, s$)}}
\label{alg:fast-nucleus}
  Enumerate all $K_r$s, mark them unprocessed, find their $\omega_s$\;
  $\ahrc \leftarrow$ list of \tsns~in a tree structure\;
  $\comp{v} = -1~~\forall~K_r~v\in G$ \tcp{\tsn~id for each $K_r$} 
  $\mcp \leftarrow$ list of adjacent \tsn~pairs\;
  \For{\meach unprocessed $K_r$ $u$ with minimum $\omega_s(u)$}{    
    $\kval(u) = \omega_s(u)$, {\tt hrc}.$max\kval=\kval(u)$\;
    $sn \leftarrow$ \tsn~with $\kval=\kval(u)$\;
    Find set $\cS$ of $K_s$s containing $u$\;
    \For{\meach $C \in \cS$}{\label{ln:fn9}
      \If{all $K_r$ $v \subset C$ is unprocessed}{
        \For{\meach $K_r$ $v \subset C$, $v \neq u$}{
    	  \lIf{$\omega_s(v) > \omega_s(u)$}{
    	    $\omega_s(v) = \omega_s(v) - 1$
    	  }
	}
      }
      \textcolor{blue}{\Else {\label{ln:fn13}
        $D \leftarrow$ set of $processed~K_r$s $\subset C$\;\label{ln:fn14}
	$w \leftarrow v \in D$ with the smallest $\kval$\;\label{ln:fn15}
	\If{$\kval(w)=\kval(u)$}{\label{ln:fn16}
          \lIf{\tcomp$(u) = -1$}{$\comp{u}=\comp{w}$}\label{ln:fn17}
   	  \lElse{\textsc{Union-r} ($\comp{u}, \comp{w}$)}\label{ln:fn18}
	} 
	\lElse {$\mcp.\tadd$($\comp{u},\comp{w}$)\tcp*[f]{$\kval(w)<\kval(u)$}}\label{ln:fn19}
      }}
    }
    \textcolor{blue}{
    \lIf{\tcomp$(u) = -1$}{{\tt comp}$(u) \leftarrow sn$, ${\tt hrc}.\tadd$($sn$)}\label{ln:fn20}
    \textnormal{Update all} $(-1,\cdot) \in \mcp$ \textnormal{with} $(\comp{u},\cdot)$\label{ln:fn21}}
  }
  \textcolor{red}{\textsc{BuildHierarchy}~($\mcp, {\tt hrc}$)\tcp*{postprocessing}\label{ln:fn22}}
  create \tsn~$root$ with $\kval=0$, tie to all existing roots\;
  ${\tt hrc}.\tadd$$(root)$ and \textnormal{\textsc{Report All the Nuclei by {\tt hrc,comp}}}\;
\end{algorithm}

\begin{algorithm}[!b]
\small
\caption{\textsc{BuildHierarchy($\mcp,{\tt hrc}$)}}
\label{alg:bh}    
    $\bmcp \leftarrow {{\tt hrc}}.max\kval$ \# of empty lists \tcp*{to bin \mcp}\label{ln:bh1}
    \For(\tcp*[f]{hrc($s$).$\kval > $~hrc($t$).$\kval$}){\meach $(s,t) \in \mcp$}{\label{ln:bh2}
      $\pbmcp{\hrc{\tn{$t$}}.\kval}.\tadd$$(s,t)$\label{ln:bh3}
    }
    \For{\meach list $l$ $\in\bmcp~in~[\ahrc.$max$\kval~...~1]$ order}{\label{ln:bh4}
      $merge \leftarrow$ \textnormal{empty list of \tsn~pairs}\label{ln:bh5}\;
      \For{\meach $(s,t) \in l$}{\label{ln:bh6}
        $s \leftarrow \textnormal{\textsc{Find-r}}(s)$, $t  \leftarrow \textnormal{\textsc{Find-r}}(t)$\label{ln:bh7}\;
        \If{$s \ne t$}{\label{ln:bh8}
          \If{$\hrc{s}.\kval > \hrc{t}.\kval$}{\label{ln:bh9}
            $\hrc{s}.\textnormal{parent} = \hrc{s}.\textnormal{root} = t$\label{ln:bh10}\;
          }
          \lElse{$merge.\tadd(s,t)$}\label{ln:bh11}
        }
      }
      \For{\meach $(s,t) \in$ merge}{\label{ln:bh12}
  	\textnormal{\textsc{Union-r}} $(s, t)$\label{ln:bh13}
      }  
    }
\end{algorithm}

\noindent All the $T_{r,s}$ can be detected without doing a traversal.
A $K_r$ is said to be processed, if its $\kval$ is assigned.
During the peeling process, neighborhood of each $K_r$ is examined, see lines~\ref{ln:sk6}-\ref{ln:sk7} of Alg.~\ref{alg:set-nucleus}, but processed neighbors are ignored (line~\ref{ln:sk8}).
We leverage those ignored neighbors to construct the $T_{r,s}$.
We introduce \textsc{FastNucleusDecomposition} algorithm (Alg.~\ref{alg:fast-nucleus}) to detect $T_{r,s}$ early in the peeling process so that costly traversal operation is not needed anymore.

\begin{table*}[!t]
\centering
\small
\linespread{0}\selectfont
\caption{\small  Statistics for the real-world graphs. Largest graph in the dataset has more than $37$M edges. $|\triangle|$ is the number of triangles, $|K_4|$ is the number of four-cliques. Ratios of $s$-cliques to $r$-cliques are shown in columns 6,7,8, for $s\le4$ and $s-r=1$.  Columns 9 to 14 show sub-$(r,s)$ nuclei numbers ($|T_{r,s}|$) and non-maximal sub-$(r,s)$ nuclei numbers ($|T^*_{r,s}|$), artifact of Alg.~\ref{alg:fast-nucleus}, for the $(r,s)$ values we interested. Last two columns are the number of connections from $T^*_{r,s}$s with higher $\lambda$ values to the ones with lower $\lambda$ values.}
\linespread{1}\selectfont
\renewcommand{\tabcolsep}{2pt}
\label{tab:properties}
\vspace*{1ex}
\begin{tabular}{|l||r|r|r|r|r|r|r|r|r|r|r|r|r|r|r|}\hline
\multicolumn{1}{|c||}{} & \multicolumn{1}{c|}{$|V|$} & \multicolumn{1}{c|}{$|E|$} & \multicolumn{1}{c|}{$|\triangle|$} &  \multicolumn{1}{c|}{$|K_4|$} & \multicolumn{1}{c|}{$\frac{|E|}{|V|}$} & \multicolumn{1}{c|}{$\frac{|\triangle|}{|E|}$} & \multicolumn{1}{c|}{$\frac{|K_4|}{|\triangle|}$} & \multicolumn{1}{c|}{$|T_{1,2}|$} & \multicolumn{1}{c|}{$|T^*_{1,2}|$} & \multicolumn{1}{c|}{$|T_{2,3}|$} & \multicolumn{1}{c|}{$|T^*_{2,3}|$} & \multicolumn{1}{c|}{$|T_{3,4}|$} & \multicolumn{1}{c|}{$|T^*_{3,4}|$} & \multicolumn{1}{c|}{$|c_{\downarrow}(T^*_{2,3})|$} & \multicolumn{1}{c|}{$|c_{\downarrow}(T^*_{3,4})|$} \\ \hline \hline

\assk	(\sassk)	&	$1.7$M	&	$11.1$M	&	$28.8$M	&	$148.8$M	&	$6.54$	&	$2.59$	&	$5.17$	&	$1.1$M	&	$	1.2	$M	&	$2.1$M	&	$	2.7	$M	&	$1.4$M	&	$	2	$M	&	$	8.0	$	M	&	$	8.2	$	M	\\ \hline
\fbbe	(\sfbbe)	&	$22.9$K	&	$852.4$K	&	$5.3$M	&	$26.6$M	&	$37.22$	&	$6.30$	&	$4.96$	&	$8.6$K	&	$	9.3	$K	&	$106.3$K	&	$	137.7	$K	&	$206.0$K	&	$	261.1	$K	&	$	1.4	$	M	&	$	1.9	$	M	\\ \hline
\fbmit	(\sfbmit)	&	$6.4$K	&	$251.2$K	&	$2.3$M	&	$13.7$M	&	$39.24$	&	$9.44$	&	$5.77$	&	$2.7$K	&	$	2.8	$K	&	$27.3$K	&	$	34.6	$K	&	$77.6$K	&	$	105.7	$K	&	$	509.1	$	K	&	$	1.3	$	M	\\ \hline
\fbst	(\sfbst)	&	$11.6$K	&	$568.3$K	&	$5.8$M	&	$37.1$M	&	$49.05$	&	$10.27$	&	$6.37$	&	$4.7$K	&	$	4.9	$K	&	$56.8$K	&	$	72.5	$K	&	$185.2$K	&	$	255.1	$K	&	$	1.2	$	M	&	$	2.9	$	M	\\ \hline
\fbtx	(\sfbtx)	&	$36.4$K	&	$1.6$M	&	$11.2$M	&	$70.7$M	&	$43.74$	&	$7.03$	&	$6.33$	&	$13.5$K	&	$	14.9	$K	&	$210.0$K	&	$	266	$K	&	$395.5$K	&	$	498.9	$K	&	$	2.6	$	M	&	$	3.7	$	M	\\ \hline
\twhg	(\stwhg)	&	$456.6$K	&	$12.5$M	&	$83.0$M	&	$429.7$M	&	$27.39$	&	$6.63$	&	$5.18$	&	$341.4$K	&	$	350	$K	&	$2.9$M	&	$	3.3	$M	&	$5.6$M	&	$	7.3	$M	&	$	27.2	$	M	&	$	59.5	$	M	\\ \hline
\wgo	(\swgo)	&	$916.4$K	&	$4.3$M	&	$13.4$M	&	$39.9$M	&	$4.71$	&	$3.10$	&	$2.98$	&	$408.7$K	&	$	508.3	$K	&	$386.9$K	&	$	568.4	$K	&	$251.4$K	&	$	382.8	$K	&	$	982.8	$	K	&	$	814.7	$	K	\\ \hline
\wuk	(\swuk)	&	$129.6$K	&	$11.7$M	&	$837.9$M	&	$52.2$B	&	$90.60$	&	$71.35$	&	$62.36$	&	$1.5$K	&	$	1.7	$K	&	$837$	&	$	837	$	&	$836$	&	$	836	$	&	$	0	$	 	&	$	0	$	 	\\ \hline
\wiki	(\swiki)	&	$3.1$M	&	$37.0$M	&	$88.8$M	&	$162.9$M	&	$11.76$	&	$2.40$	&	$1.83$	&	$2.4$M	&	$	2.5	$M	&	$7.9$M	&	$	9.7	$M	&	$6.7$M	&	$	9.3	$M	&	$	45.1	$	M	&	$	45.0	$	M	\\ \hline
\end{tabular}
\vspace{-3ex}
\end{table*}

At each iteration of the peeling process, a $K_r$ with the minimum $\omega$ is selected and $\omega$ of its unprocessed neighbors are decremented.
No information about the surrounding processed $K_r$s is used.
If we check the processed neighbors, we can infer some connectivity information and use it towards constructing all the $T_{r,s}$ as well as the hierarchy-skeleton.
For example, assume we are doing $k$-core decomposition and a vertex $u$ with degree $d$ is selected.
We assign $\kval(u)=d$ and check the unprocessed neighbors of $u$ to decrement their degree, if greater than $d$.
We can also examine the processed neighbors.
$\kval$ of any processed neighbor is guaranteed to be less than or equal to $d$, by definition.
Say $v$ is a neighbor with $\kval(v)=d$.
Then, we can say that $u$ and $v$ are in the same $T_{r,s}$.
Say $w$ is another neighbor with $\kval(w)<d$.
Then, we can infer that maximum $d$-$(1,2)$ nucleus of $w$ contains $u$, and $T_{r,s}$ of $u$ is an ancestor of $T_{r,s}$ of $w$ in the hierarchy-skeleton.
Leveraging these pairwise relations enables us to find all the $T_{r,s}$ and construct the hierarchy-skeleton.

An important thing to note is that, it is not always possible to detect the $T_{r,s}$ of a $K_r$ by only looking at the processed neighbors.
Consider $k$-core decomposition on a star graph, for which all vertices has $\kval=1$.
Center vertex is processed in the last two steps of peeling, so it is not possible to infer two connected vertices with equal $\kval$ until that time.
We find \emph{non-maximal $T_{r,s}$}s (denoted as $T^*_{r,s}$) and combine them by using the disjoint-set forest algorithm.
The difference from the \textsc{DF-Traversal} algorithm is that our hierarchy-skeleton will have more nodes because of non-maximal $T_{r,s}$.

Colored lines~\ref{ln:fn13}-\ref{ln:fn22} in Alg.~\ref{alg:fast-nucleus} implements our ideas.
For each $K_s$ we encountered (line~\ref{ln:fn9}), processed neighbors are explored starting from line~\ref{ln:fn13}.
Note that, there is no need to check every adjacent and processed $K_r$ in the same $K_s$, since the relations among them are already checked in previous steps.
It is enough to find and process the $K_r$ $w$ with minimum $\kval$, as in line~\ref{ln:fn15}.
If $w$ has an equal $\kval$ value (line~\ref{ln:fn16}), we need to either put our $K_r$ of interest to the \tsn~of $w$ (line~\ref{ln:fn17}) or merge to the \tsn~of $w$ by \textsc{Union-r} operation (line~\ref{ln:fn18}).
In \textsc{FastNucleusDecomposition} algorithm, we only build disjoint-set forests during the peeling process (until line~\ref{ln:fn22}).
If $\kval(w)$ happens to have a smaller value (line~\ref{ln:fn19}), we put the pair of \tsns~to a list (\mcp), which will be used to build the hierarchy-skeleton after the peeling.
We do not process the relations between \tsns~of different $\kval$ right away for two reasons: (1) \tsn~of the $K_r$ of interest might not be assigned yet (\comp{$u$} is -1), (2) order of processing \tsns~relations is crucial to build the hierarchy-skeleton correctly and efficiently.
Regarding (1), we take care of the $K_r$s not belonging to a \tsn~in lines~\ref{ln:fn20} and~\ref{ln:fn21}.
For (2), we have the \textsc{BuildHierarchy} function (line~\ref{ln:fn22}), defined in Alg.~\ref{alg:bh}.

In \textsc{BuildHierarchy}, we create $max\kval$ number of bins to distribute the \tsns~pairs based on the smaller $\kval$ of the \tsn~pair.
The reason is same with our reverse order discovery of \tsns~in \textsc{DF-Traversal} (Alg.~\ref{alg:DFtraverse-nucleus}): we construct the hierarchy-skeleton in a bottom-up manner and it enables us to use disjoint-set forest algorithm to locate the $k$-$(r,s)$ nuclei that we need process.
Distribution is done in lines~\ref{ln:bh2}-\ref{ln:bh3}.
Then, we just process the binned list (\bmcp) in reverse order of $\kval$ values (line~\ref{ln:bh4}).
We do the same operations to build the hierarchy-skeleton: lines~\ref{ln:bh7}-\ref{ln:bh11} of \textsc{BuildHierarchy} and lines~\ref{ln:sn18}-\ref{ln:sn22} of \textsc{SubNucleus} algorithm (Alg.~\ref{alg:subnucleus}) are almost same.
Once we finish each list in \bmcp, we union the accumulated \tsns~of equal $\kval$ values (lines~\ref{ln:bh12}-\ref{ln:bh13}), as we did in lines~\ref{ln:sn24}-\ref{ln:sn25} of \textsc{SubNucleus} algorithm.
Finally, in \textsc{FastNucleusDecomposition} we create a \tsn~ to represent entire graph, make it {\tt parent} to all parentless \tsns, and report the hierarchy.

Avoiding traversal does not change the time complexity of overall algorithm, since the peeling part was already taking more time.
Auxiliary data structures in \textsc{FastNucleusDecomposition} requires additional space, though.
\ahrc~needs $4\cdot|T^*_{r,s}|$ in which \tsns~are not necessarily maximal, and {\tt comp} needs $|K_r|$.
\mcp~structure corresponds to the connections from $T^*_{r,s}$s with higher $\lambda$ values to the ones with lower $\lambda$ values, which we denote as $c_{\downarrow}(T^*_{r,s})$.
The upper bound for $|c_{\downarrow}(T^*_{r,s})|$ is ${s \choose r}|K_s|$, when each $K_r$ is assumed to be a $T^*_{r,s}$ and their $\lambda$ values are adversary (see the end of Section~\ref{subsec:gener} for details).
However, it is quite unlikely as we show in Section~\ref{sec:exp}.
\bmcp~in \textsc{BuildHierarchy} is just an ordered version of \mcp, and needs the same amount of space; $|c_{\downarrow}(T^*_{r,s})|$.
Lastly, \textit{merge} in \textsc{BuildHierarchy} might require another $|c_{\downarrow}(T^*_{r,s})|$ at most, but it is quite unlikely.
Overall, additional space requirement of \textsc{FastNucleusDecomposition} at any instant is $4\cdot|T^*_{r,s}|+2\cdot |c_{\downarrow}(T^*_{r,s})|+|K_r|$ and additional $|c_{\downarrow}(T^*_{r,s})|$ might be needed.
Section~\ref{sec:exp} gives more details on $T^*_{r,s}$ structure of real-world networks and their impact to the memory cost.

\section{Experiments}\label{sec:exp}
\noindent We evaluated our algorithms on different types of real-world networks, obtained from SNAP~\cite{snap}, Network Repository~\cite{nr} and UF Sparse Matrix Collection~\cite{UFL}.
Our dataset includes an internet topology network (\assk), facebook friendship networks of some universities (\fbbe, \fbmit, \fbst, \fbtx)~\cite{Traud12}, follower network of Twitter users tweeted about Higgs boson-like particle discovery
 (\twhg), web networks (\wgo, \wuk) and network of wikipedia pages (\wiki).
We ignore the directions for directed graphs.
Important statistics of the networks are given in Table~\ref{tab:properties}.
$|K_s|$/$|K_r|$ ratio gives an estimate for the $k$-$(r,s)$ nucleus decomposition runtime, as explained at the end of Section~\ref{subsec:gener}, and we put them in columns 6-8 to show the challenging and diverse characteristics of the networks in our dataset.
Note that most of the networks have relatively high edge density in the realm of real-world networks and it makes the computation more expensive.
We also included graphs with various $|K_s|$/$|K_r|$ ratios to diversify our dataset.
Last eight columns are shown to explain the runtime and memory costs of our algorithms.
All the algorithms are implemented in {\tt C++} and compiled using {\tt gcc 5.2.0} at {\tt -O2} optimization level.
All experiments are done on a Linux operating system running on a machine with Intel Xeon Haswell E5-2698 2.30 GHz processor with 128 GB of RAM.

Nucleus decomposition has been shown to give denser subgraphs and more detailed hierarchies for $s-r$=1 cases, for fixed $s$~\cite{Sariyuce15}.
We implemented and tested our algorithms for $s-r$=1 cases where $s\le4$: giving us $(1,2), (2,3)$ and $(3,4)$ nucleus decompositions.
$(1,2)$-nucleus decomposition is same as the $k$-core decomposition~\cite{Seidman83} and $(2,3)$ corresponds to $k$-truss community finding~\cite{Huang14} (stronger definition of $k$-truss decomposition~\cite{Cohen08}).
We consider the $k$-$(r,s)$ nucleus as a set of $K_r$s.
In our algorithms, we find the $k$-$(r,s)$ nuclei for all $k$ values and determine the hierarchy tree among those nuclei.
We report the total time of peeling and traversal (or post-processing) that takes the graph as input and gives all the nuclei with an hierarchy.

\subsection{ k-core decomposition (or (1, 2) nuclei)}

\begin{table}[!t]
\vspace*{-2ex}
\centering
\small
\linespread{0}\selectfont{}
\caption{\small $k$-core decomposition results. \textnormal{\textsc{Hypo}} is the hypothetical limit for the best possible traversal based algorithm. \textnormal{\textsc{Naive}}~is Alg.~\ref{alg:naive}, \textnormal{\textsc{DFT}} is the one using Alg.~\ref{alg:DFtraverse-nucleus}, \textnormal{\textsc{FND}} is the Alg.~\ref{alg:fast-nucleus} and \textnormal{\textsc{LCPS}} is our adaptation from~\protect\cite{MaBe83}. Right-most column is the runtimes of fastest algorithm and rest are its speedups over other algorithms.}
\linespread{1}\selectfont{}
\renewcommand{\tabcolsep}{2pt}
\vspace*{1ex}
\begin{tabular}{|l||r|r|r|r|r|r|r|}\hline
& \multicolumn{4}{c|}{speedups with respect to} & time (s) \\
\multicolumn{1}{|c||}{$\textbf{(1,2)}$\textbf{-nuclei}} & \multicolumn{1}{c|}{\textsc{Hypo}} &  \multicolumn{1}{c|}{\textsc{Naive}} & \multicolumn{1}{c|}{\textsc{DFT}} & \multicolumn{1}{c|}{\textsc{FND}} & \multicolumn{1}{c|}{\textsc{LCPS}} \\ \hline \hline
\assk	&$	0.41	$x&$	5.94	$x&$	2.79	$x&$	1.33	$x&$	1.94	$\\ \hline
\fbbe	&$	0.82	$x&$	19.74	$x&$	1.68	$x&$	2.77	$x&$	0.05	$\\ \hline
\fbmit	&$	0.82	$x&$	17.51	$x&$	1.74	$x&$	2.80	$x&$	0.01	$\\ \hline
\fbst	&$	0.78	$x&$	25.50	$x&$	1.76	$x&$	2.19	$x&$	0.03	$\\ \hline
\fbtx	&$	0.78	$x&$	23.01	$x&$	1.61	$x&$	2.57	$x&$	0.11	$\\ \hline
\twhg	&$	0.80	$x&$	27.89	$x&$	1.94	$x&$	2.60	$x&$	1.18	$\\ \hline
\wgo	&$	0.15	$x&$	3.45	$x&$	0.40	$x&$	0.27	$x&$	3.83	$\\ \hline
\wuk	&$	0.86	$x&$	58.02	$x&$	2.61	$x&$	3.32	$x&$	0.17	$\\ \hline
\wiki	&$	0.52	$x&$	10.09	$x&$	1.96	$x&$	1.41	$x&$	7.79	$\\ \hline
\textbf{avg}	& $\textbf{0.66}$\textbf{x} & $\textbf{21.24}$\textbf{x} & $\textbf{1.83}$\textbf{x} & $\textbf{2.14}$\textbf{x} & \textbf{best} \\ \hline
\end{tabular}
\label{tab:12res}
\end{table}

\noindent Peeling phase of the $k$-core decomposition, which finds $\kval$ values of vertices, is well studied, and most efficient  implementation is used in our work.
As mentioned before, traversal part, however, is mostly overlooked and there is no true algorithm to serve as a baseline.

In their pioneering work, Matula and Beck~\cite{MaBe83} introduced a high-level algorithm, named \textsc{LCPS} (Level Component Priority Search), to detect $k$-cores and the hierarchy among them.
\textsc{LCPS} algorithm traverses the vertices based on $\kval$ values in a selective order and outputs them with interspersed brackets.
The vertices enclosed by paired brackets at depth $k+1$ are the vertices of a $k$-core in G.
The traversal can be started from any vertex, and neighbors are discovered and put in a priority queue along with their $\kval$ values.
At each step, vertex with the maximum $\kval$ is chosen from the queue, and processed.
Closed/open brackets are interspersed according to the $\kval$ values of consecutively processed vertices.
Authors argue that time complexity of \textsc{LCPS} is $O(|E|)$, but \emph{an implementation may not always be possible owing to the difficulty of maintaining an appropriate priority queue}~\cite{MaBe83}.

\begin{table*}
\centering
\small
\linespread{0}\selectfont{}
\caption{\small $\textbf{(2,3)}$ and $\textbf{(3,4)}$ nucleus decomposition results. Right-most column for each is the runtimes of the fastest algorithm and rest are its speedup over other algorithms. Starred numbers (*) show lower bounds, when the algorithm did not finish in 2 days. \textnormal{\textsc{Hypo}} is the hypothetical limit for the best possible traversal based algorithm. For $\textbf{(2,3)}$, it is done on edges and triangle connections, and for $\textbf{(3,4)}$ it is on triangles and four-clique connections. \textnormal{\textsc{Naive}} is Alg.~\ref{alg:naive} and \textnormal{\textsc{TCP*}} is the indexing algorithm proposed by~\protect\cite{Huang14} and does not even include $(2,3)$ nuclei finding. \textnormal{\textsc{DFT}} is the one using Alg.~\ref{alg:DFtraverse-nucleus} and \textnormal{\textsc{FND}} is the Alg.~\ref{alg:fast-nucleus}.}
\linespread{1}\selectfont{}
\renewcommand{\tabcolsep}{2pt}
\begin{tabular}{|l||r|r|r|r|r||r|r|r|r|r|}\hline
\multicolumn{1}{|c||}{} & \multicolumn{5}{c||}{$\textbf{(2,3)}$\textbf{-nuclei}} & \multicolumn{4}{c|}{$\textbf{(3,4)}$\textbf{-nuclei}}\\
& \multicolumn{4}{c|}{speedups with respect to} & \multicolumn{1}{c||}{time (s)} & \multicolumn{3}{c|}{speedups with respect to} & \multicolumn{1}{c|}{time (s)} \\ \hline
\multicolumn{1}{|c||}{} & \multicolumn{1}{c|}{\textsc{Hypo}} &  \multicolumn{1}{c|}{\textsc{Naive}} & \multicolumn{1}{c|}{\textsc{TCP*\tiny{\protect\cite{Huang14}}}} & \multicolumn{1}{c|}{\textsc{DFT}} & \multicolumn{1}{c||}{\textsc{FND}} & \multicolumn{1}{c|}{\textsc{Hypo}} &  \multicolumn{1}{c|}{\textsc{Naive}} & \multicolumn{1}{c|}{\textsc{DFT}} & \multicolumn{1}{c|}{\textsc{FND}}  \\ \hline \hline
\assk	&$	1.54	$x&$	45.62	$x&$	3.69	$x&$	2.04	$x&$	91.3	$&$	1.78	$x&$	163.91^*$x&$	1.96	$x&$	1054.2	$\\ \hline
\fbbe	&$	1.13	$x&$	10.76	$x&$	3.41	$x&$	1.40	$x&$	7.3	$&$	1.42	$x&$	1812.47^*$x&$	1.52	$x&$	95.3	$\\ \hline
\fbmit	&$	1.10	$x&$	11.68	$x&$	3.45	$x&$	1.33	$x&$	2.8	$&$	1.45	$x&$	3848.02^*$x&$	1.53	$x&$	44.9	$\\ \hline
\fbst	&$	1.10	$x&$	12.58	$x&$	3.41	$x&$	1.36	$x&$	7.8	$&$	1.48	$x&$	1321.89^*$x&$	1.58	$x&$	130.7	$\\ \hline
\fbtx	&$	1.10	$x&$	14.41	$x&$	3.35	$x&$	1.37	$x&$	16.8	$&$	1.49	$x&$	679.06^*$x&$	1.57	$x&$	254.5	$\\ \hline
\twhg	&$	1.33	$x&$	16.24	$x&$	3.27	$x&$	1.49	$x&$	255.5	$&$	1.78	$x&$	38.96^*$x&$	1.81	$x&$	4434.9	$\\ \hline
\wgo	&$	1.31	$x&$	1729.93	$x&$	3.90	$x&$	1.59	$x&$	13.0	$&$	1.35	$x&$	1083.02^*$x&$	1.43	$x&$	159.6	$\\ \hline
\wuk	&$	1.68	$x&$	90.50	$x&$	11.07	$x&$	3.64	$x&$	562.5	$&$	1.24	$x&$	1.98^*$x&$	1.98^*	$x&$	87329.6	$\\ \hline
\wiki	&$	1.53	$x&$	7.06	$x&$	3.37	$x&$	1.63	$x&$	584.9	$&$	1.82	$x&$	23.00^*$x&$	1.91	$x&$	7513.5	$\\ \hline
\textbf{avg}	&$  	\textbf{1.31}	$\textbf{x}& $	\textbf{215.4}	$\textbf{x} & $	 \textbf{4.32}	$\textbf{x} & $	\textbf{1.76}	$\textbf{x} & $	\textbf{best}	$&$	\textbf{1.53}	$\textbf{x} & $>	\textbf{996.92}$\textbf{x} & $	>\textbf{1.70}$	\textbf{x} & $	 \textbf{best} 	$\\ \hline
\end{tabular}
\label{tab:234res}
\vspace*{-4ex}
\end{table*}

We adapt and implement the \textsc{LCPS} algorithm for our objectives.
We alleviate the problem of \emph{maintaining an appropriate priority queue} by using the bucket data structure, known by the bucket sort~\cite{CLRS01}.
During the traversal, we place the discovered vertices in the bucket according to their $\kval$ values, and find the one with the maximum $\kval$ value in $O(1)$ time.
We also adapted the placing brackets part to have the tree hierarchy, where each node is a $T_{1,2}$.
If the $\kval$ value of vertex of interest is equal to the one in the previous step, we stay in the same node.
If it is greater, we go down in the tree by creating a chain of new nodes, otherwise we climb up in the tree.
Number of these new nodes (or steps to climb) is the difference between $\kval$s of current and previous vertex.
As a result, we create a tree hierarchy with as many levels as the max $\kval$ in the graph.

Table~\ref{tab:12res} gives the total $k$-core decomposition runtimes for experimented algorithms; the fastest one shown on the right-most column and its speedups over other algorithms given on the corresponding columns.
\textsc{Naive} is the base nucleus decomposition algorithm, given in Alg.~\ref{alg:naive}, and uses the naive traversal to find all $k$-cores.
\textsc{DFT} is the algorithm that uses \textsc{DF-Traversal} (Alg.~\ref{alg:DFtraverse-nucleus}) and \textsc{FND} is the \textsc{FastNucleusDecomposition} algorithm (Alg.~\ref{alg:fast-nucleus}).
\textsc{LCPS} denotes the runtime of our LCPS adaptation.
Note that we are not aware of any other adaptation/implementation of the \textsc{LCPS} algorithm in~\cite{MaBe83}.
We also have a hypothetical runtime, shown as \textsc{Hypo}, which is the peeling time plus a BFS traversal on the entire graph.
\textsc{Hypo} is a limit for the most efficient traversal (plus regular peeling) that can be done on a given graph.
Although it does not compute the hierarchy and find $k$-cores, it shows the best that can be done with a traversal-based nucleus decomposition algorithm.
Peeling phases of \textsc{Hypo}, \textsc{Naive}, \textsc{DFT}, and \textsc{LCPS} are same.

Results show that our \textsc{LCPS} adaptation outperforms other alternatives.
Detailed timings suggest that traversal time is almost same with peeling, satisfying the linear complexity in the number of edges.
On average, \textsc{LCPS} is $21.24$x faster than \textsc{Naive} algorithm.
Speedups over \textsc{DFT} and \textsc{FND} are $1.83$x and $2.14$x.
Our \textsc{LCPS} adaptation also runs in only $66$\% time of the theoretical limit of \textsc{Hypo}, where the overhead is due to the bucket data structure.

\subsection{(2, 3) nuclei}
\noindent Table~\ref{tab:234res} gives the runtime comparison for entire $(2,3)$ nucleus decomposition, where rightmost column is the fastest algorithm and other columns are its speedups over other algorithms.
To start with, we state that our peeling implementation for $(2,3)$ nucleus decomposition is quite efficient; for instance~\cite{WaCh12} computes $\kval$ values of \assk~graph in 281 secs where we can do it in 74 secs.
We also tested {\tt wiki-Talk} graph just to compare with~\cite{Huang14}, and we get $41/13=3.15$ times faster results despite our less powerful CPU.
Apart from \textsc{Naive}, \textsc{DFT} and \textsc{FND} algorithms, we give \textsc{Hypo} as the bound for the best possible traversal-based nucleus decomposition algorithm that traverses graph over edges and triangles.
In addition, we implemented the \textsc{TCP} index construction algorithm introduced by Huang \emph{et al.}~\cite{Huang14}.
In that work, authors devise \textsc{TCP} index on vertices which enable fast traversal to answer max-$(2,3)$ nucleus queries on vertices.
\textsc{TCP} index is a tree structure at each vertex, which is actually the maximum spanning forest of each ego-network.
\textsc{TCP} index is constructed after the peeling process to be ready for incoming queries.
Note that our implementation of \textsc{TCP} index construction is also quite efficient that is $1.8$x faster than~\cite{Huang14} for the dataset in their work.
In Table~\ref{tab:234res}, we show the time only for peeling plus \textsc{TCP} index construction.
Entire graph still needs to be traversed using \textsc{TCP} to find all the $(2,3)$ nuclei and the hierarchy among them.
Peeling phases of  \textsc{Hypo}, \textsc{Naive}, \textsc{TCP} and \textsc{DFT} are same.

\begin{figure}[!b]
\centering
 \includegraphics[width=0.99\linewidth]{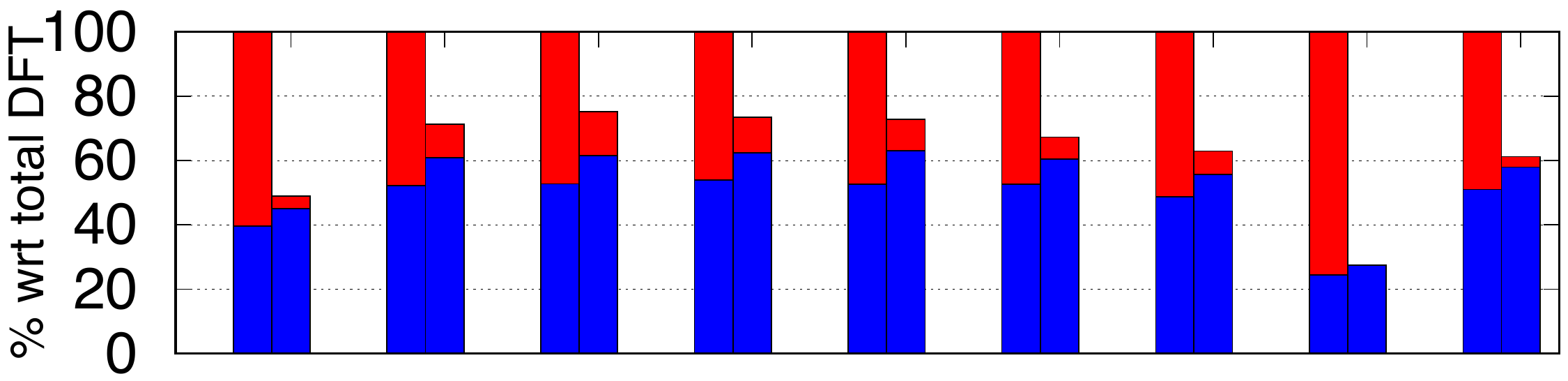}
 \vspace{-1ex}
 \includegraphics[width=0.99\linewidth]{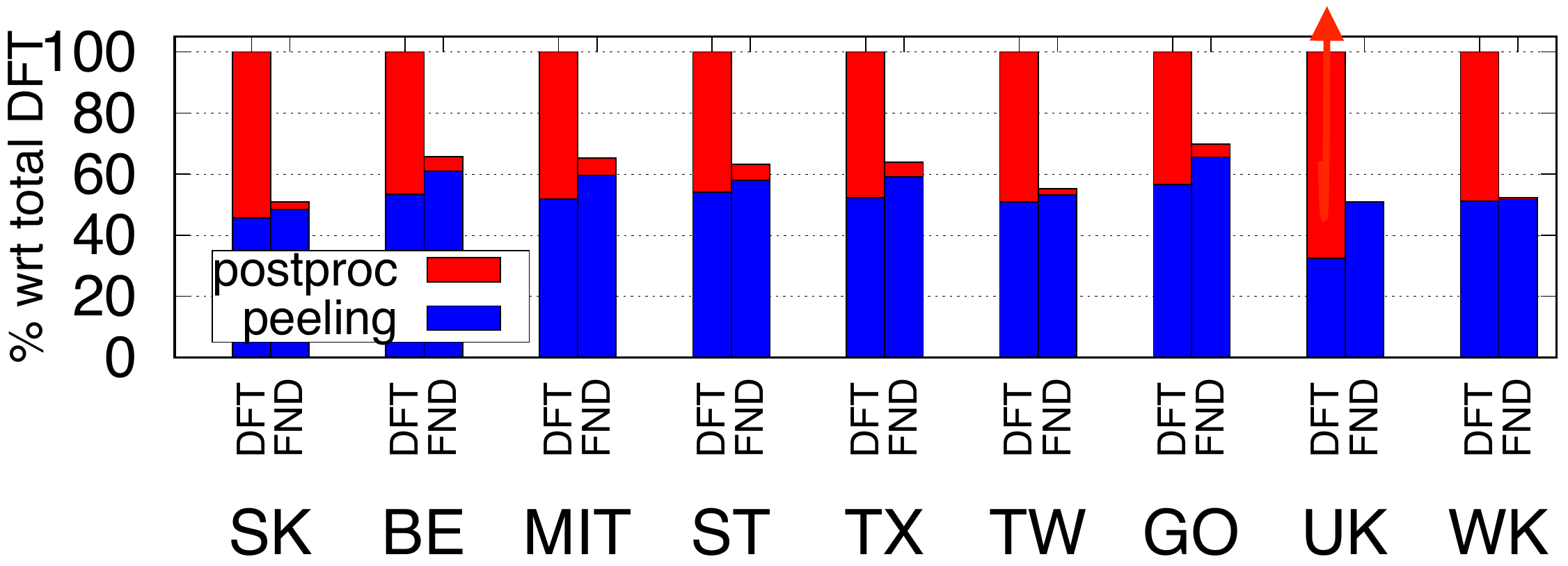}
\vspace*{-1ex}
\linespread{0}\selectfont{}
\caption{\small $\textbf{(2,3)}$ [top] and $\textbf{(3,4)}$ [bottom] nucleus decomposition comparison for \textsc{DFT} (the one using Alg.~\ref{alg:DFtraverse-nucleus} for traversal) and \textsc{FND} (shown in Alg.\ref{alg:fast-nucleus}). Two main results: (1) Traversal part of \textsc{DFT} is close to the peeling part, (2) \textsc{FND} is able to keep the total runtime comparable to the peeling part of \textsc{DFT}}
\linespread{1}\selectfont{}
\label{fig:234ch} 
\vspace*{-1ex}
\end{figure}

Before checking the fastest, we look at the \textsc{DFT} algorithm, which is the second best.
It is $2.4$x and $122.4$x faster than \textsc{TCP} and \textsc{Naive} algorithms, respectively.
Note that, \textsc{TCP} time is \emph{even before the traversal}.
In $40$\% of the time that \textsc{TCP} spends for being ready to answer queries, \textsc{DFT} is able to give all the answers, i.e, $(2,3)$ nuclei and the hierarchy.
This shows the benefit of using disjoint-set forests to detect the nuclei and construct the hierarchy.
One of our initial objectives was to meet the computational complexity of traversal phase by true design and implementation.
\textsc{DFT} does that by keeping the traversal time close to the peeling time; traversal takes only $23$\% more time than peeling on average.
Figure~\ref{fig:234ch} shows the comparison, where we normalize runtimes with respect to total \textsc{DFT}.
Also, \textsc{DFT} performs only $34$\% slower than the hypothetical bound (\textsc{Hypo}).

The fastest algorithm is \textsc{FND}.
It is $215$x faster than \textsc{Naive} on average and gives $1729$ speedup on \wgo.
It outperforms \textsc{TCP} by $4.3$x on average and up to $11$x (on \wuk).
\textsc{FND} also works $1.76$x faster than \textsc{DFT} and Figure~\ref{fig:234ch} shows the detailed comparison.
For \textsc{DFT}, peeling is Alg.~\ref{alg:set-nucleus} and postprocessing is the traversal time (Alg.~\ref{alg:DFtraverse-nucleus}).
For \textsc{FND}, postprocessing is the \textsc{BuildHierarchy} algorithm (Alg.~\ref{alg:bh}) and peeling is the rest in Alg.~\ref{alg:fast-nucleus}.
\textsc{FND} is able to keep the total time comparable to the peeling of \textsc{DFT}; it is only $29$\% slower!
This is exactly the benefit of avoiding traversal by early discovery of non-maximal $T_{2,3}$.
On \wuk~graph, benefit of avoiding traversal is most apparent; \textsc{DFT} performs worst compared to \textsc{FND}.
Because there are only 837 $T_{2,3}$s which means no work is done by disjoint-set forest and all the time is spent on traversal (see Figure~\ref{fig:234ch}).
As mentioned in Section~\ref{sec:alg}, extended peeling in \textsc{FND} finds non-maximal $T_{2,3}$, and the count might be much larger than the number of $T_{2,3}$s.
However, it is not the case.
Last 6 columns of Table~\ref{tab:properties} show that there is no significant difference; on average non-maximal $T_{2,3}$s are $24$\% more than $T_{2,3}$s.

\textsc{DFT} and \textsc{FND} algorithms require additional memory, as mentioned at the end of Section~\ref{subsec:dstraversal}.
The upper bound of additional space for \textsc{DFT} is $6\cdot|T_{2,3}|+3\cdot|E|$, and takes at most $\sim$650MB for any graph in our dataset, where {\tt int} is used to store each number (four bytes).
Although the upper bound of $|T_{2,3}|$ is $|E|$, we observed that it is only $13.1$\% of $|E|$ on average, which can be calculated from Table~\ref{tab:properties}.
Regarding the \textsc{FND}, upper bound is given as $4\cdot|T^*_{2,3}|+3\cdot |c_{\downarrow}(T^*_{2,3})|+|E|$ (see Section~\ref{subsec:ditchtraversal}) and it takes at most $\sim$1.4GB for the experimented graphs (two {\tt int}s for each connection in $c_{\downarrow}(T^*_{2,3})$).
We observed that \wuk~graph have so many edges with no triangles that results in isolated $T^*_{2,3}$s (and $T^*_{3,4}$s), so $|c_{\downarrow}(T^*_{2,3})|=|c_{\downarrow}(T^*_{3,4})|=0$.
For the rest, we see that $|c_{\downarrow}(T^*_{2,3})|$ is only $8.7$\% of the upper bound (${3 \choose 2}|\triangle|$), on average.

Lastly, we check \textsc{FND} vs. \textsc{Hypo}.
\textsc{FND} is faster on all instances, and $1.31$x on average, meaning that it outperforms \textbf{any possible} traversal-based algorithm for $(2,3)$ nucleus decomposition.
This is so important to show the strength of (1) detecting the non-maximal $T_{2,3}$ early to avoid traversal, (2) using disjoint-set forest data structure to handle on-the-fly computation.

\vspace{-1ex}
\subsection{(3, 4) nuclei}
\noindent Table~\ref{tab:234res} and Figure~\ref{fig:234ch} gives the comparison of all algorithms on $(3,4)$-nucleus decomposition, which has been shown to give subgraphs with highest density and most detailed hierarchy~\cite{Sariyuce15}.
Results are similar to the $(2,3)$ case: \textsc{FND} is the best and speedups are sharper.
\textsc{Naive} algorithm could not be completed in 2 days for any graph.
\textsc{DFT} also could not finish its computation on \wuk~graph in that time.
Regarding the memory requirement, we see that \textsc{DFT} and \textsc{FND} take at most $\sim$10GB and $\sim$3.5GB for the graphs in our dataset.
Also, $|T_{3,4}|$ and $|c_{\downarrow}(T^*_{3,4})|$ are far from the upper bounds, only of $3.9$\% and $2.5$\% on average.
Overall, \textsc{FND} outperforms all others.
Figure~\ref{fig:234ch} shows that total time of \textsc{FND} is only $21$\% more than the peeling time of \textsc{DFT} and only $2$\% slower on \wiki.
Most significantly, \textsc{FND} is $1.53$x faster than hypothetical limit (\textsc{Hypo}) of any possible traversal-based $(3,4)$ nucleus decomposition algorithm.

\vspace{-1ex}
\section{Conclusion}
\noindent In this work, we focused on computing the hierarchy of dense subgraphs given by peeling algorithms.
We first provided a detailed review of previous work and pointed misconceptions about $k$-core and $k$-truss decompositions.
We proposed two generic algorithms to compute \emph{any} nucleus decomposition.
Our idea to leverage disjoint-set forest data structure for hierarchy computation works well in practice.
We further improved the performance by detecting the subgraphs during peeling process to avoid traversal.
We also adapted and implemented an existing idea for $k$-core decomposition hierarchy, and showed its benefit.
Overall, our algorithms significantly outperformed the existing alternatives.

There are two open questions that might be worth to look at.
First is about the analysis.
Nested structures given by the resulting hierarchy only show the $k$-$(r,s)$ nuclei.
Instead looking at the $T_{r,s}$s, which are many more than the $k$-$(r,s)$ nuclei, might reveal more insight about networks.
This actually corresponds to the hierarchy-skeleton structure that our algorithms produce.
Second is regarding the performance.
We believe that adapting the existing parallel peeling algorithms for the hierarchy computation can be helpful.\\
\vspace{-2ex}
\section{Acknowledgments}
This work was funded by the Applied Mathematics Program at the U.S. Department of Energy and Sandia's Laboratory Directed Research \& Development (LDRD) program.
Sandia National Laboratories is a multi-program laboratory managed and operated by Sandia Corporation, a wholly owned subsidiary of Lockheed Martin Corporation,
for the U.S. Department of Energy's National Nuclear Security Administration under contract DE-AC04-94AL85000.
\vspace{-1ex}

\balance

\bibliographystyle{abbrv}
\renewcommand{\baselinestretch}{0.6}
\scriptsize
\bibliography{paper}

\end{document}